\documentclass[aps,prl,onecolumn,superscriptaddress,showpacs]{revtex4}

\usepackage{graphics}
\usepackage{bm}\usepackage{amsmath}    
\usepackage{amssymb}    
\usepackage{graphicx}   
\usepackage{subfigure}

\newcommand{\jun}{junction }

\begin{document}

\title {$\delta$-biased Josephson tunnel junctions}
\thanks{Submitted to Phys. Rev. B.}

\author{R. Monaco}
\affiliation{Istituto di Cibernetica del CNR, 80078, Pozzuoli, Italy
and Unit$\grave{\rm a}$ INFM $­$ Dipartimento di Fisica, Universit$\grave{\rm a}$ di Salerno, 84081 Baronissi, Italy}\email
{roberto@sa.infn.it}
\author{J. Mygind}
\affiliation{DTU Physics, B309, Technical University of
Denmark, DK-2800 Lyngby, Denmark}
\author{V.\ P.\ Koshelets and P. Dmitriev}
\affiliation{Institute of Radio Engineering and Electronics,
Russian Academy of Science, Mokhovaya 11, Bldg 7, 125009, Moscow, Russia.}

\date{\today}
\begin{abstract}
The behavior of a long Josephson tunnel junction drastically depends
on the distribution of the dc bias current. We investigate the case
in which the bias current is fed in the central point of a
one-dimensional junction. Such junction configuration has been
recently used to detect the persistent currents circulating in a
superconducting loop. Analytical and numerical results indicate that
the presence of fractional vortices leads to remarkable differences
from the conventional case of uniformly distributed dc bias current.
The theoretical findings are supported by detailed measurements on
a number of $\delta$-biased samples having different electrical and
geometrical parameters.
\end{abstract}


\maketitle

\section{  Introduction}

\begin{figure}[t]
\centering
\subfigure[ ]{\includegraphics[width=5cm]{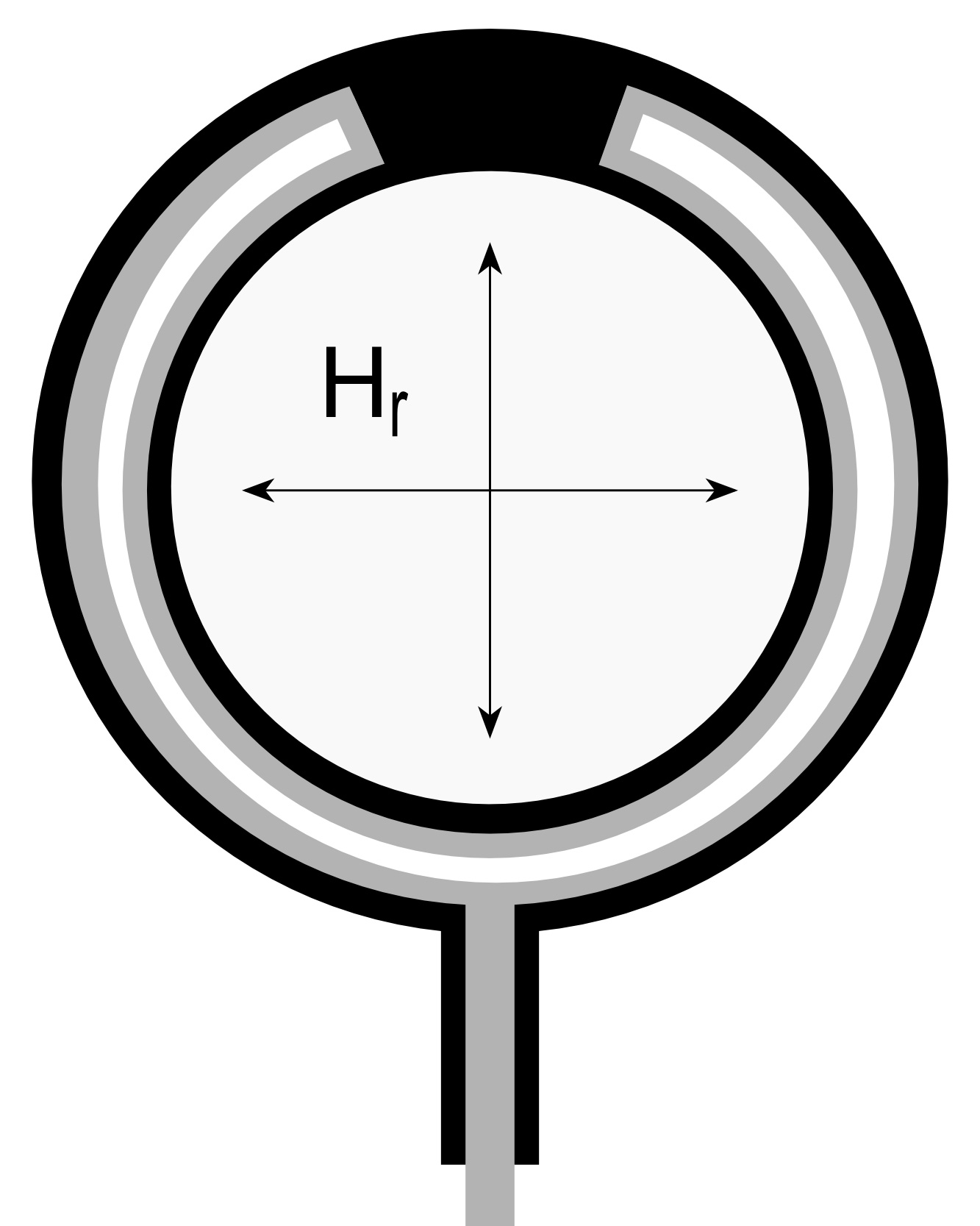}}
\subfigure[ ]{\includegraphics[width=7cm]{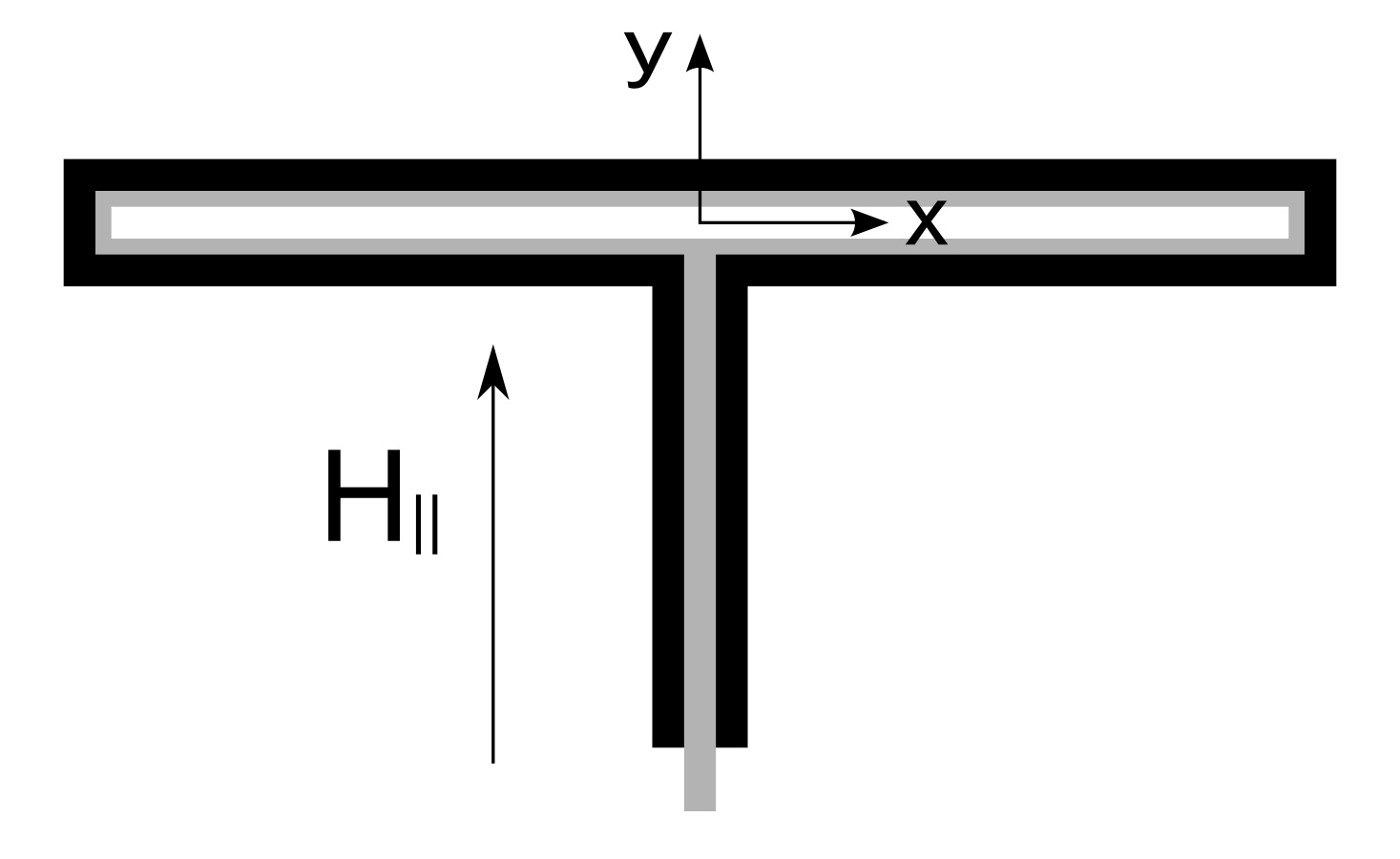}}
\caption{(a) gapped annular and (b) linear $\delta$-biased Josephson tunnel junctions. The base electrodes are in black, the top electrodes are in gray and the junction areas are white. A gapped annular junction in a radially uniform magnetic field $H_r$ is topologically equivalent to a linear junction in an in-plane uniform field $H_{||}$.}
\label{2D}
\end{figure}

Symmetry principles and how they are broken are fundamental concepts
in physics. A recent experiment\cite{PRB09,viewpoint} demonstrated
that the symmetry can spontaneously break during the fast
superconducting phase transition of a metal ring and both fluxoids
or antifluxoids can be trapped in the ring while its temperature
crosses the superconducting critical temperature. This phenomenon
was predicted a long time ago as one among several possible
condensed matter \textit{cosmological} experiments\cite{zurek2} to
check the validity of the \textit{causality} principle in the early
Universe\cite{kibble1}. In the experiment of Ref.\cite{PRB09} the
presence of the persistent currents associated with the flux trapped
in the ring relied on the radial magnetic field modulation of the
critical current of a planar Josephson tunnel junction (JTJ) having the peculiar configuration
shown in Fig.~\ref{2D}(a). It consists of a ring shaped junction
\textit{cut} at some point with the bias current fed at the
diametrally opposite point. The cut leaves a \textit{gapped} annular
junction and reliefs the junction from the constraint of the
$2\pi$-periodic boundary condition of annular
junctions\cite{PRB96}. In that experiment the ring itself acted as
the junction base electrode, while the top electrode had the shape
of a circular arc of about $300^\circ$. Later on it will be
demonstrated that a gapped annular junction in a radially uniform
magnetic field $H_r$ is topologically equivalent to a linear
junction in an in-plane uniform field $H_{||}$ as that depicted in
Fig.~\ref{2D}(b).

\noindent The task of this work is to study the properties of a
\textit{$\delta$-biased} or \textit{single point injected}
 overlap Josephson tunnel junction whose physical length is $L$, i.e., $-L/2\leq X \leq L/2$ and whose width is  $W$, i.e.,
$-W/2\leq Y \leq W/2$. To simplify the analysis, we assume that
Josephson current density $J_c$ is uniform over the barrier area and
that the junction width $W$ is smaller than the Josephson
penetration depth $\lambda _{J}$. Ideally, the bias current is fed to the junction by infinitely narrow electrodes. In real devices the
$\delta$-bias approximation is achieved as far as the electrodes
carrying the bias current in and out of the tunnel barrier are much
narrower than the junction Josephson penetration depth $\lambda_j$;
however, for very long junctions, it is only required that the
electrodes widths are much smaller than the junction length $L$. In
window type JTJs one more requirement is that the passive region
surrounding the tunnel area, the so-called \textit{idle
region}\cite{lee,JAP95,Ustinov}, needs to be narrower than the
current-carrying electrodes otherwise the bias current diffuses
before entering the barrier and the sharp bias profile gets smeared.
We will consider both intermediate length ($L\simeq \lambda_j$) and
long ($L>>\lambda_j$) JTJs (the behavior of small junctions is not
affected by the bias profile). We remark that with the current
injected at the junction extremities we recover the well-known case
of so called \textit{in-line} configuration treated by pioneering
works on long JTJs soon after the discovery of the Josephson effect
\cite{ferrel,stuehm,OS}. For in-line JTJs it is important to
distinguish between the symmetric\cite{stuehm,OS} and asymmetric
configuration\cite{ferrel,stuehm}: the former is achieved when the
bias current enters at one extremity and exits at the opposite one,
while the latter is obtained when the bias current enters and exits
from the same extremity. Since in this paper we will only consider
the case in which the bias current is fed in the middle of
the junction long dimension, for symmetry reasons, we do not need to
specify the electrode configuration. Nevertheless the more general
situation in which the bias current enters and leaves in two generic
lateral points along the junction will deserve consideration in the
future. We will show that, with the bias current centrally injected,
both the static and dynamic junction properties reveal interesting
phenomena whose understanding will serve as a base for the study of
more general cases. In this regard, a considerable attention has
been recently given to the case of $0-\pi$ transition Josephson
tunnel junction obtained by closely situated current
injectors\cite{goldobin04,goldobin05}. For the sake of completeness
it is worth to remark that the lateral point injected bias was
already treated in the literature, but only in the limit of
infinitely long JTJs\cite{kuprianov,likharev,radparvar}.

\section{  The model}

\begin{figure}[htb]
        \centering
                \includegraphics[width=9cm]{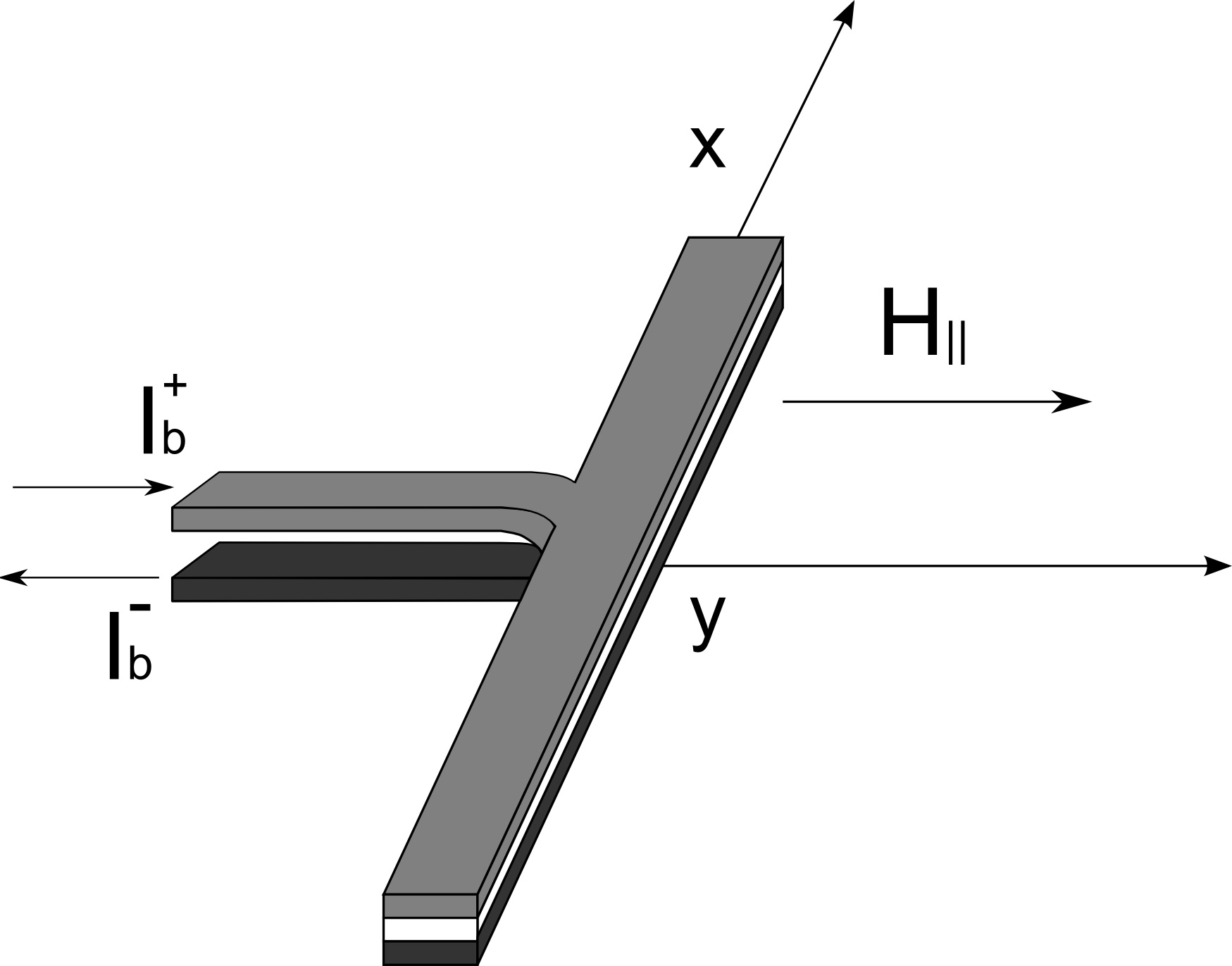}
        \caption{ 3D sketch of a centrally injected $\delta$-biased linear planar Josephson tunnel junction.}
        \label{3D}
\end{figure}

It can be shown that for a $\delta$-biased JTJ, as that sketched in
Fig.~\ref{3D}, the gauge-invariant phase difference $\phi$ of the
order parameters of the superconductors on each side of the tunnel
barrier obeys the static or d.c. perturbed sine-Gordon equation:

\begin{equation}
\lambda_j^2 \frac{d^2 \phi}{d X^2} = \sin \phi(X) - \gamma \delta \left( \frac{X}{\lambda_j} \right),
\nonumber
\label{sG}
\end{equation}

\noindent in which the term $\gamma=I_b/I_0$ is the external bias
current $I_b$ normalized to $I_0=J_cLW$ and $\delta$ is the
$\delta$-function. In normalized units of $x= X/\lambda_j$, the
above partial differential equation (PDE) becomes:

\begin{equation}
\frac{d^2 \phi}{d x^2} = \sin \phi(x) - \gamma \delta(x). \label{PDE}
\end{equation}

\noindent If the time $t$ replaces $x$ in Eq.(\ref{PDE}) and a new phase
$\psi=\phi-\pi$ is defined, we obtain the equation for a lossless
pendulum with an applied torque like a train of pulses with unitary
period. In this framework, the boundary conditions will become
constraints on the pendulum angular velocity half period before and
after each pulse.

\vskip 5pt

\noindent It is possible to derive that, introducing the  Heaviside
step function ${\mathcal{H}}(x)$, the Lagrangian density
$\mathcal{L}$ of our system is constant:

\begin{equation}
\mathcal{L}(x)= \frac{\phi_x^2 }{2}+ \cos \phi + \gamma \phi_x(0) {\mathcal{H}}(x).
\nonumber
\end{equation}

\noindent By subtracting the constant quantity $\gamma \phi_x(0)/2$, we can rewrite the last expression as:

\begin{equation}
\frac{\phi_x^2 }{2}+ \cos \phi + \frac{\gamma \phi_x(0)}{2}  \textrm{sgn}(x)= C,
\label{energy}
\end{equation}

\noindent with $C$ being a constant depending on both the external bias and magnetic field.  Note that if $\phi_x(x)$ is
discontinuous in $x=0$, then $2 \phi_x(0)= \phi_x(0+)+\phi_x(0-)$.
The first two terms in Eq.(\ref{energy}) are related to the system
free energy density $\mathcal{E}(x)=1-\cos \phi - \phi_x^2/2$, while
the last term represents the two-level potential $\mathcal{U}(x)$
generated by the $\delta$-shaped forcing term.

\subsection{  Boundary conditions}

The magnetic Josephson equation\cite{joseph} states that the phase gradient is proportional to the magnetic field:

\begin{equation}
\label{gra}
{\bf \nabla} \phi = \frac{2\pi d_e \mu_0}{\Phi_0}{\bf H}\times {\bf n} ,
\end{equation}

\noindent where ${\bf n}$ is a unit vector normal to the insulating
barrier separating the two superconducting electrodes. If the two superconducting films have thicknesses $d_{1,2}$
and London penetration depths $\lambda_{L1,2}$ and $t_j$ is the
barrier thickness, then the effective magnetic penetration $d_e$ is
given by\cite{wei}:

\begin{equation}
d_e=t_j + \lambda_{L1} \tanh \frac{d_1}{2 \lambda_{L1}} + \lambda_{L2} \tanh \frac{d_2}{2 \lambda_{L2}},
\label{d_e}
\nonumber
\end{equation}

\noindent which, in the case of thick superconducting films ($d_i
>> \lambda_{Li}$), reduces to $d_e \approx \lambda_{L1} +
\lambda_{L2}$ (since always $d_i >> t_j$).

\noindent From Eq.(\ref{gra}) it follows that, for a linear JTJ in
an external uniform field $H_{||}$ applied in the junction plane perpendicular to its length $L$, the boundary
conditions are:

\begin{equation}
\left. \frac{d \phi}{d X} \right|_{X=\pm L/2} =\kappa H_{||}, \label{bc}
\end{equation}

\noindent with $\kappa  ={2\pi d_e \mu _0}/{\Phi_0}$. Being $\lambda^{-2}_{j}=2\pi d_{e} \mu _{0} J_{c}/\Phi_0=\kappa J_c$ ($\Phi_0$ is the magnetic flux quantum and $\mu_0$ is the
vacuum permeability), we have $\kappa \lambda_j= 1/J_c \lambda_j$.
Introducing the critical field $H^*_c$ for a unitary junction with a
Fraunhofer magnetic diffraction pattern $H^*_c=\Phi_0/(\mu_0 d_e
\lambda_j)= 2\pi J_c \lambda_j=2\pi/\kappa \lambda_j$, we get $\kappa \lambda_j = 2\pi/
H^*_c$. In normalized units of  $h=\kappa \lambda_j H_{||}=2\pi
H_{||}/H^*_c$, the boundary conditions (\ref{bc}) for Eq.(\ref{PDE}) are:

\begin{equation}
\left. \frac{d \phi}{d x} \right|_{x=\pm l/2}=h,
\label{bcn}
\end{equation}

\noindent in which we have introduced the junction normalized length
$l=L/\lambda_j$. Note that, with this notations, the normalized
critical field $h^*_c$ of a short JTJ equals $2\pi/l$. For the
gapped annular junction in a radially uniform field $H_r$, the
boundary conditions for the PDE in Eq.(\ref{PDE}) are independent on
the gap angle\cite{martucciello} and coincide with those in
Eq.(\ref{bcn}), but now with $h=2\pi H_r/H^*_c$. In other words, a linear
junction in an in-plane field $H_{||}$ and a gapped annular junction
in a radial field $H_r$ are governed by the same PDE with the same boundary conditions. Therefore, for
the remaining of the paper we will use the properly normalized field
$h$ both for linear and gapped annular JTJs. A radial magnetic field
can be generated by a current flowing in a control line in the shape
of  a loop concentric to the annulus; however, the simplest
way\cite{PRB09} is to have a ring shaped base electrode and to apply
an external field perpendicular to it to induce tangential screening
currents proportional to the applied field. The only disadvantage of
this method is that the effective radial field felt by the gapped
annular junction depends on geometrical factors such as the ring
inner and outer radii and the junction position relative to the
ring. The radial dependence of the current circulating in a
superconducting ring has been calculated under many different
conditions\cite{brandt}.

\vskip 5pt

\noindent From Eqs.(\ref{PDE}) and (\ref{bcn}), we have that the
junction Josephson current $i_j$ is proportional to the bias
current; in fact, being $\int_{-a}^{a}\delta(x)dx=1$, we get:

\begin{equation}
i_j \equiv \frac{I_j}{I_0}=\frac{1}{l} \int_{-l/2}^{l/2}\sin \phi(x) dx=  \frac{1}{l} \left( \left. \frac{d \phi}{d x} \right|_{x=+l/2}-\left. \frac{d \phi}{d x} \right|_{x=-l/2} \right) + \frac{\gamma}{l} = \frac{\gamma}{l},
\label{lineare}
\end{equation}

\noindent  that is, for a fixed bias current $I_b$, the zero-voltage
current $I_j$ passing through a $\delta$-biased JTJ is inversely
proportional to its normalized length ($I_j= I_b\lambda_j/L$). In
the well-known case of an overlap JTJ with uniform bias
$\gamma(x)=\gamma_u$, it would be $i_j=\gamma_u$ (i.e., $I_j=I_b$),
meaning that $\gamma_u$ cannot exceed unity. In contrast,
Eq.(\ref{lineare}) implies that for $\delta$-biased junctions the
largest value $\gamma_c$ that the normalized bias current can
achieve is determined by the junction normalized length. Later on it
will be found that, in our case, $|\gamma| \leq 4$.

\vskip 5pt

\noindent  The jump in the phase gradient at the injection point
(the axis origin in  case) can be calculated directly from
Eq.(\ref{PDE}). For any $0<x_0<l/2$ we can write:

$$ \left. \frac{d \phi}{d x} \right|_{x=x_0}- \left. \frac{d \phi}{d x} \right|_{x=-x_0} =  \int_{-x_0}^{x_0} [\sin \phi(x) - \gamma \delta(x) ] dx= \int_{-x_0}^{x_0}  \sin \phi(x) dx - \gamma. $$

\noindent Taking the limit $x_0 \to 0$, the integral vanishes, being
$\phi(x$) a continuous function; then the phase derivative jumps at the injection point $x=0$:

\begin{equation}
 \left. \frac{d \phi}{d x} \right|_{x=0+}- \left. \frac{d \phi}{d x} \right|_{x=0-} = - \gamma. \label{jump}
\end{equation}

\noindent This equation was first reported by Kuprianov \textit{et
al.}\cite{kuprianov} [see also Ref.\cite{likharev} at (par.8.5)] for
JTJs with lateral single-point injection. We can rewrite
Eq.(\ref{jump}) as:

\begin{equation}
 \left. \frac{d
\phi}{d x} \right|_{x=0-}=- \left. \frac{d \phi}{d x} \right|_{x=0+}
= \gamma/2+ h_0,
\label{bc1}
\end{equation}

\noindent in which the constant $h_0$ is a measure of the
(exponential) penetration of the external magnetic field (if any) into the center of
the junction and, in general, cannot be determined {\it a priori}.

\section{  Analytical approximate solutions for long JTJs}

In this paragraph we will discuss the possible analytical approaches
to determine the phase profile of a point injected long JTJ in the
Meissner state. In absence of external bias and magnetic field the
Josephson phase profile $\phi(x)$ is identically equal to zero. If
$\gamma$ and $h$ are sufficiently small, then $|\phi(x)|<<1$, so
that Eq.(\ref{PDE}) can be linearized, being $\sin \phi \approx
\phi$. Barone \textit{et al.}\cite{vaglio,barone} pointed out that a
piecewise linear current phase relationship $j_j=\chi^2 \phi \leq 1$
(with $\chi \in [0.5,1]$) can correctly handle the case when
$|\phi(x)|<\pi/2$. Obviously in a linear approximation the phase profile looses its
correct structure within the scale of $\lambda_j$, but averaged
results such as the Josephson current are not significantly affected
by the approximation. However, as will be shown later on, for
$\delta$-biased JTJs in the Meissner state it is $|\phi(x)|\leq \pi$
so that also the piecewise linear approximation fails and a cubic
approximation can be conveniently adopted $\sin \phi \simeq \alpha
\phi - \beta \phi^3$ leading to a Duffing-like differential
equation for the oscillation of a soft ($\alpha, \beta >0$) spring
system. In the special case of the underdamped and unforced Duffing
equation, exact solutions can be written in terms of Jacobi's
elliptic functions\cite{tamura}. Further, approximate solutions of
the forced Duffing equation could be found using the perturbation
methods\cite{jordan} when $\beta<<\alpha$. However, simple solutions
can be readily obtained in the approximation of very long junctions.

\subsection{  $h=0$ and $\gamma \neq 0$, $l/2 > 2\pi$}

\begin{figure}[tb]
\centering
\subfigure[ ]{\includegraphics[width=7cm]{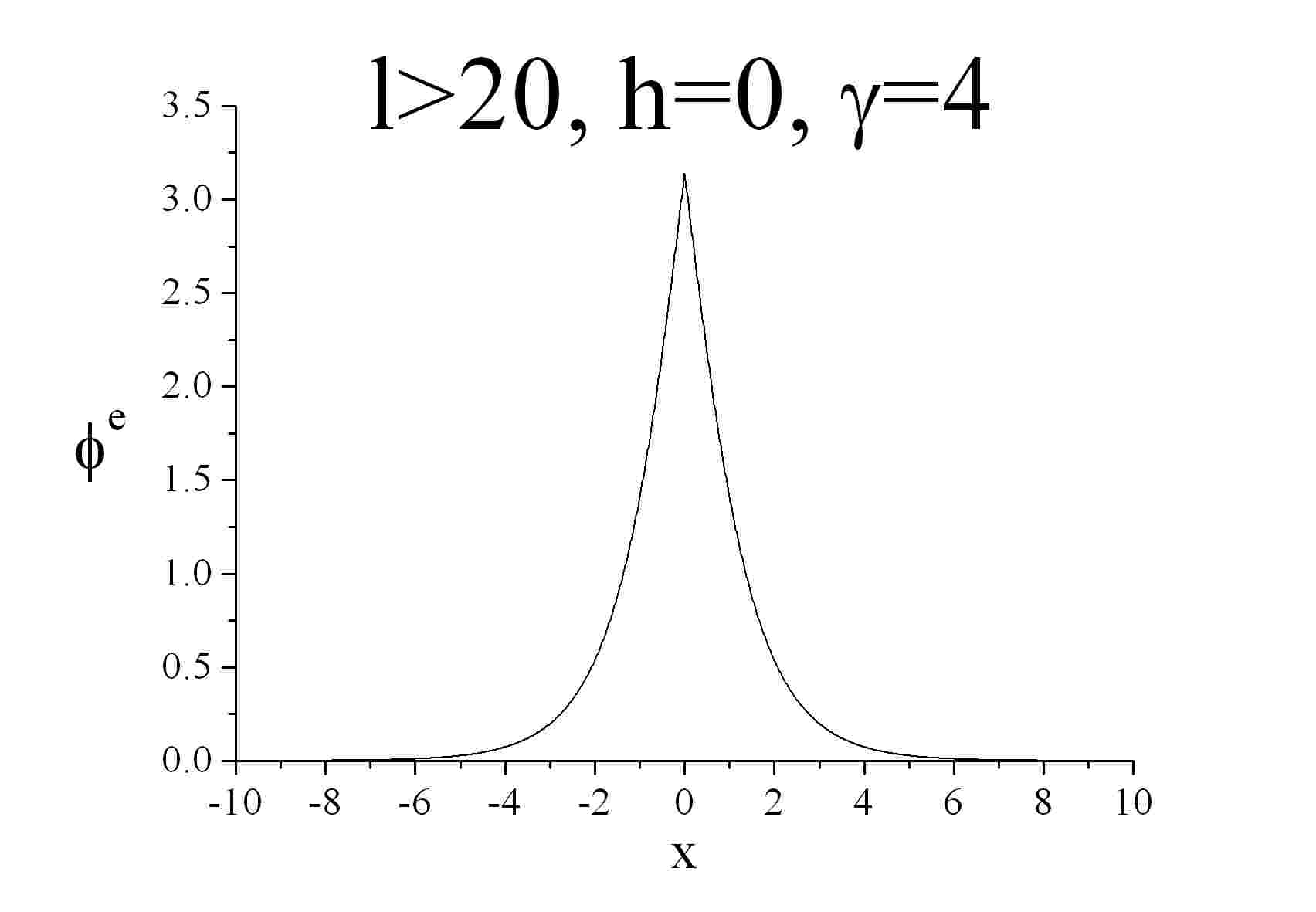}}
\subfigure[ ]{\includegraphics[width=7cm]{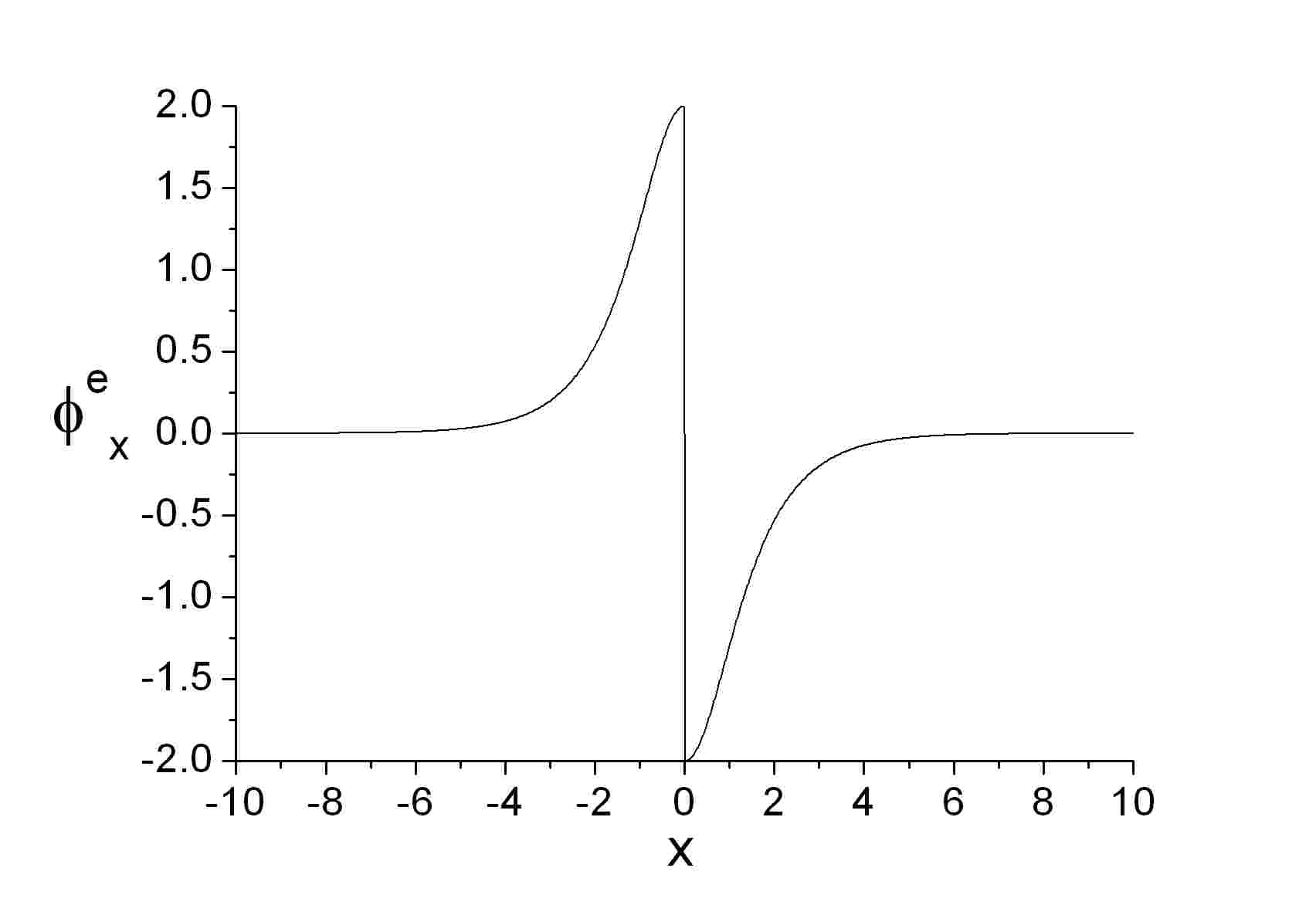}}
\subfigure[ ]{\includegraphics[width=7cm]{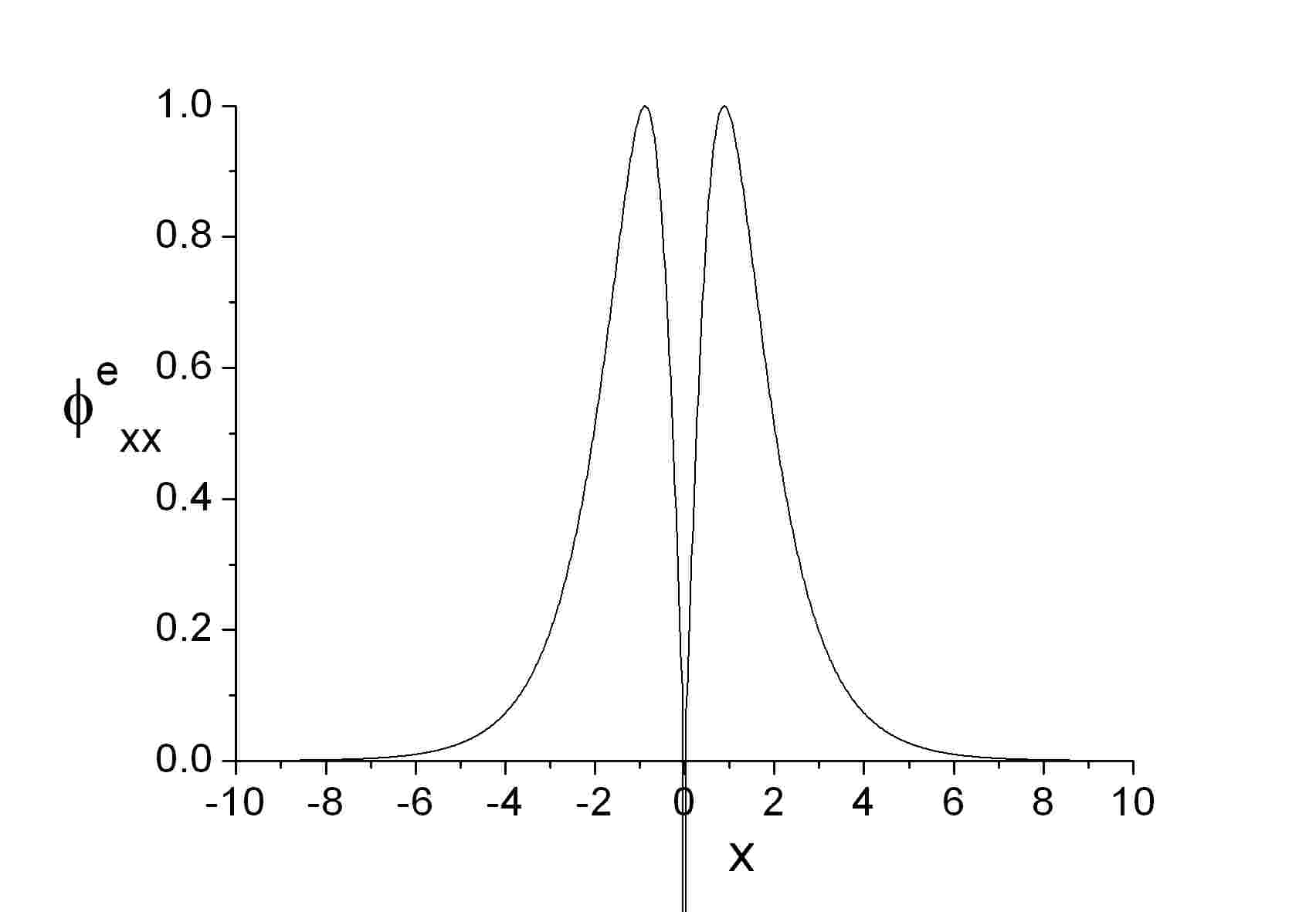}}
\subfigure[ ]{\includegraphics[width=7cm]{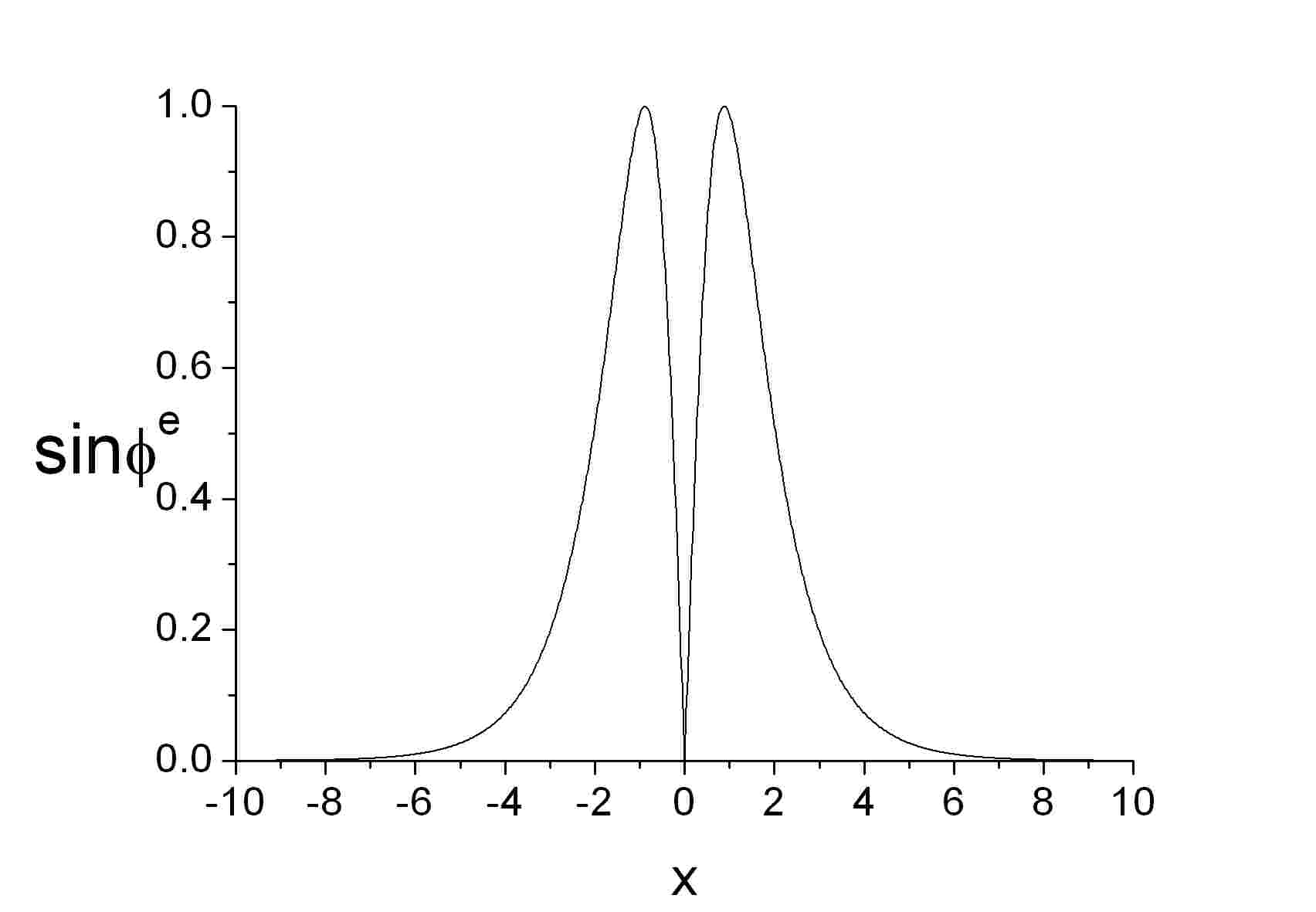}}
\caption{(a) Phase profile $\phi^e(x)$ for an infinitely long $\delta$-biased JTJ as in (\ref{phie}) with $\xi=0$, for $x \in [-10,10]$. (b), (c) and (d) show, respectively,  $\phi_x^e(x)$, $\phi_{xx}^e(x)$, and $\sin \phi^e(x)$. It is $\cos \phi^e(x)= 1-[\phi^e_{x}(x)]^2/2$.}
\label{figPhie}
\end{figure}

\noindent We begin with the case of no externally applied magnetic
field which, in the well-studied case of uniform bias, results in a
constant phase profile $\phi(x)= \sin^{-1}\gamma_u$ (mod $2\pi$).
For symmetry reasons, in the case of $\delta$-biased JTJs, the
solution of Eq.(\ref{PDE}) has to be an even function $\phi^e(x)$.
Therefore its gradient is an odd function  such that
$\phi_x^e(x)+\phi_x^e(-x)=0$ for any $x \in ]0,l/2]$, meaning that
$h_0=0$ in Eq.(\ref{bc1}). With $h_0=0$, Eq.(\ref{bc1}) provides two
extra conditions on the phase left and right derivative at the
injection point:

\begin{equation}
\left. \frac{d
\phi}{d x} \right|_{x=0-}=- \left. \frac{d \phi}{d x} \right|_{x=0+}
= \gamma/2. \label{bc0}
\end{equation}

\noindent For $l>>4\pi$ (strictly in the limit $l \to \infty$) with
the above conditions, an approximate solution of Eq.(\ref{PDE}) is a
cusp-like function:

\begin{equation}
\phi^e(x)=  \pm 4 \tan^{-1} \exp -(\left|x\right|+\xi),
\label{phie}
\end{equation}

\noindent in which $\xi$ is a non-negative constant set by the bias
current $\gamma$; Eq.(\ref{phie}) can also be cast in the
form\cite{ferrel}: $\sin
\phi^e(x)/2=\pm \textrm{sech}(\left|x\right|+\xi)$. It is $\phi^e(\pm
\infty)=\phi_x^e(\pm \infty)=0$ and $\phi^e(0) = \pm 4 \tan^{-1}\exp
(-\xi)$. The sign in front of Eq.(\ref{phie}) concords with that of
the external bias current. With $\xi=0$ (and positive $\gamma$),
then $\phi^e(0)=\pi$, meaning that Eq.(\ref{phie}) corresponds to a
semifluxon ($\pi$ jump) on the left side of the junction and anti
semifluxon ($-\pi$ jump) on the right side, as shown in
Figs.~\ref{figPhie}(a-d) for $-10<x<10$ (we observe that, as
required, $\phi_{xx} = \sin \phi$, everywhere, but for $x=0$). The
dependence of $\xi$ on $\gamma$ can be found observing that, in
force of Eq.(\ref{jump}), $\gamma= \phi_x^e(0-)-\phi_x^e(0+) =\pm
4\textrm{sech} \xi$, so the largest possible amplitude value for the
normalized bias $|\gamma_c|=4$ is achieved when $\xi=0$. For $\xi<0$
the phase in the origin grows above the threshold value
$|\phi^e(0)|=\pi$ and the static solution in Eq.(\ref{phie}) is no
longer stable: the two semifluxons develop into integer fluxons
driven away in opposite directions under the effect of the Lorentz
force associated with the bias current. According to
Ref.\cite{likharev} and from numerical simulations (reported later),
we found that the instability arises when the amplitude of $\gamma$
exceeds the critical value $\gamma_c=4$, corresponding to $Ic=\pm 4
J_c \lambda_j W$. With such a notation $\gamma=\pm \gamma_c
\textrm{sech}\xi$, i.e., $\xi = \cosh^{-1} (\gamma_c/|\gamma|)$. The
last expression allows us to find the dependence of $\phi^e(0)$ on
$\gamma$; it is found that $\cos \phi^e(0)=1-2(\gamma/\gamma_c)^2$
(i.e., $ \sin \phi^e(0)/2= \gamma/\gamma_c$). We remark that,
inserting the value $\gamma_c=4$ in Eq.(\ref{lineare}), we reach the
important conclusion that for very long junctions the normalized
zero-field critical current is inversely proportional to the the
junction length:

\begin{equation}
i_c(h=0)=  \frac{\gamma_c}{l},
\label{ic}
\end{equation}

\noindent as for asymmetric inline junctions\cite{ferrel} with the
only difference that $\gamma_c=2$. Indeed, a long $\delta$-biased
JTJ in zero external field is equivalent to two inline asymmetric
junctions in a parallel configuration; this will not any longer be
true in presence of an external field. For a generic $\gamma$ value
in the interval $[-4,4]$ the phase difference $\Delta \phi=
\phi(0)-\phi(-\infty)=\phi(0)=2 \sin^{-1}\gamma/\gamma_c$
corresponds to what in nowadays language is called a $k$-fractional
vortex where $k=\Delta \phi /(2 \pi)=\pi^{-1}
\sin^{-1}\gamma/\gamma_c $. Semi ($\xi=0$) and fractional ($\xi>0$)
vortices are presently receiving a great deal of attention in the
context of $0-\pi$ transition Josephson
junctions\cite{kirtley,lazarides,goldobin02,goldobin04}.

\vskip 5pt

\noindent In summary, the main difference between short and long
$\delta$-biased JTJs is that in the former case the solution becomes
unstable when $\gamma>\gamma_c=l$ and $\phi^e(0)>\pi/2$, while in
the latter case the phase profile becomes unstable when
$\gamma>\gamma_c=4$ and $\phi^e(0)>\pi$. The gradual crossover from
intermediate to long JTJs has to be calculated numerically; it was
found to be nicely described by the following empirically found
relationships:

\begin{equation}
\gamma_c(h=0,l)=4 \tanh \frac{l}{4},
\label{gammac}
\end{equation}

\begin{equation}
\phi_c^e(x=0,h=0,l)=\frac{\pi}{2}+ \tan^{-1} \exp (l- \pi).
\label{phic}
\end{equation}

\subsection{ $h \neq 0$ and $\gamma = 0$, $l/2>2\pi$ }

\noindent With $\gamma=0$, the nonlinear PDE Eq.(\ref{PDE}) reduces to:

\begin{equation}
\frac{d^2 \phi}{d x^2} = \sin \phi(x).
\label{OS}
\end{equation}

\begin{figure}[tb]
\centering
\subfigure[ ]{\includegraphics[width=7cm]{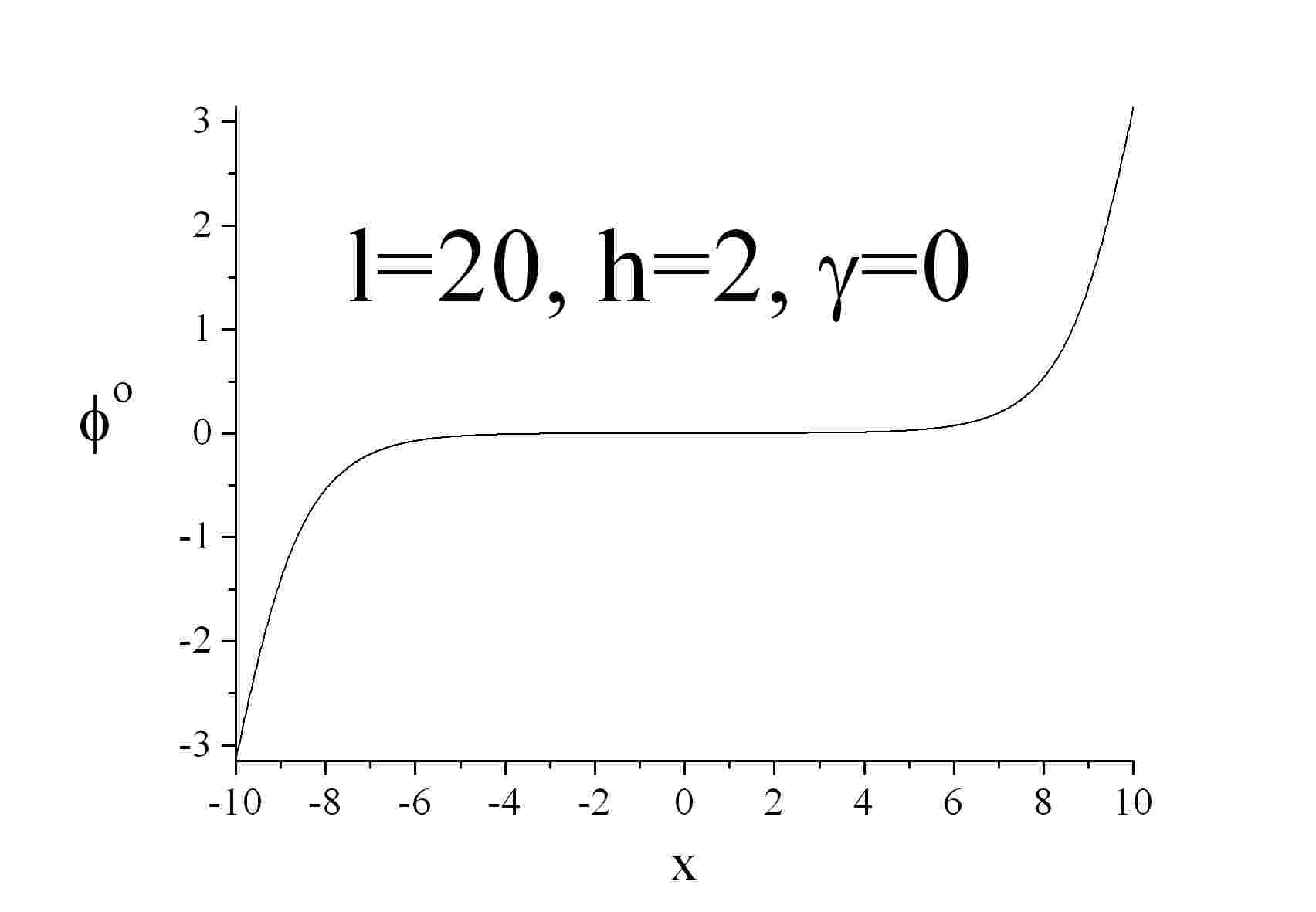}}
\subfigure[ ]{\includegraphics[width=7cm]{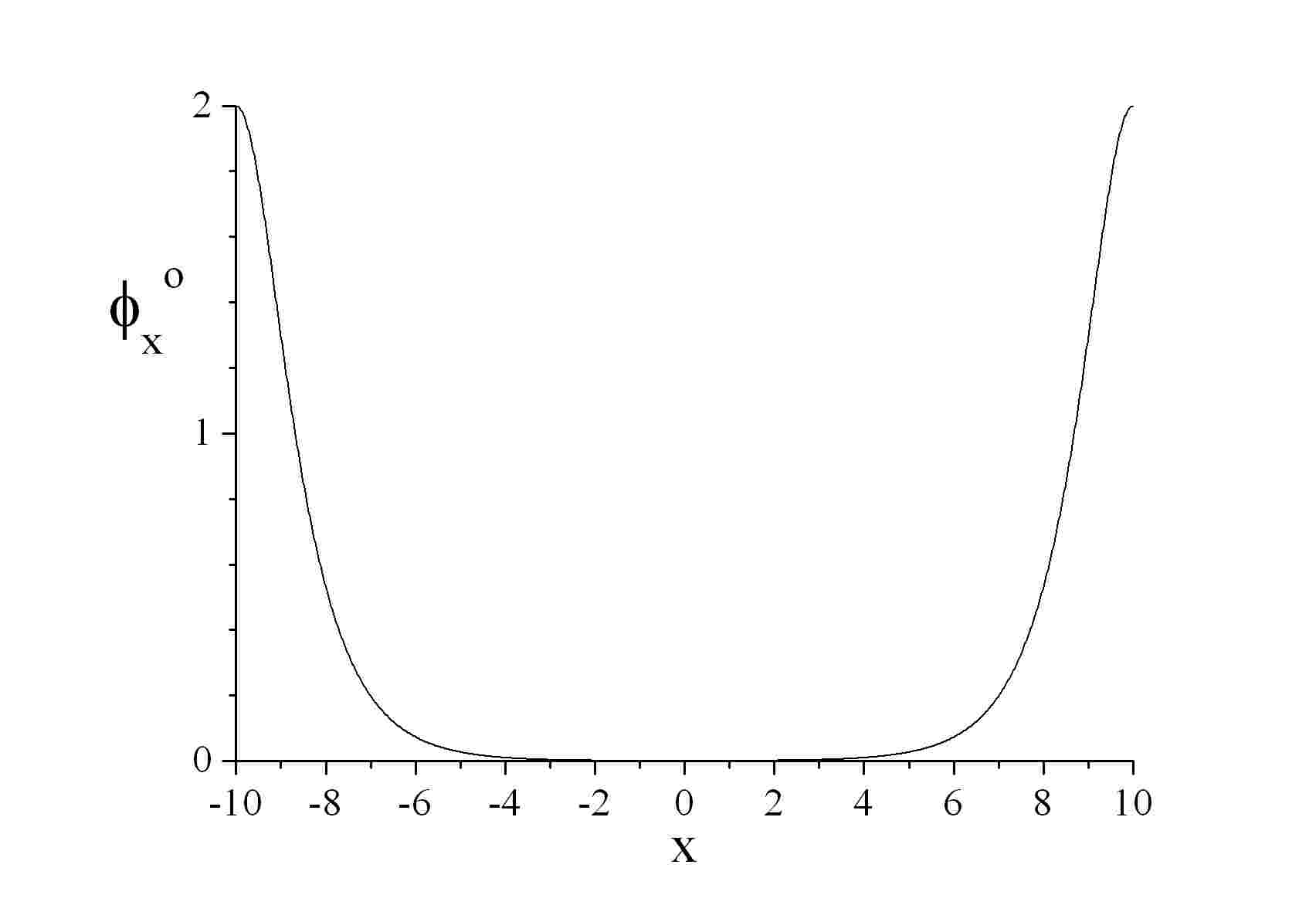}}
\subfigure[ ]{\includegraphics[width=7cm]{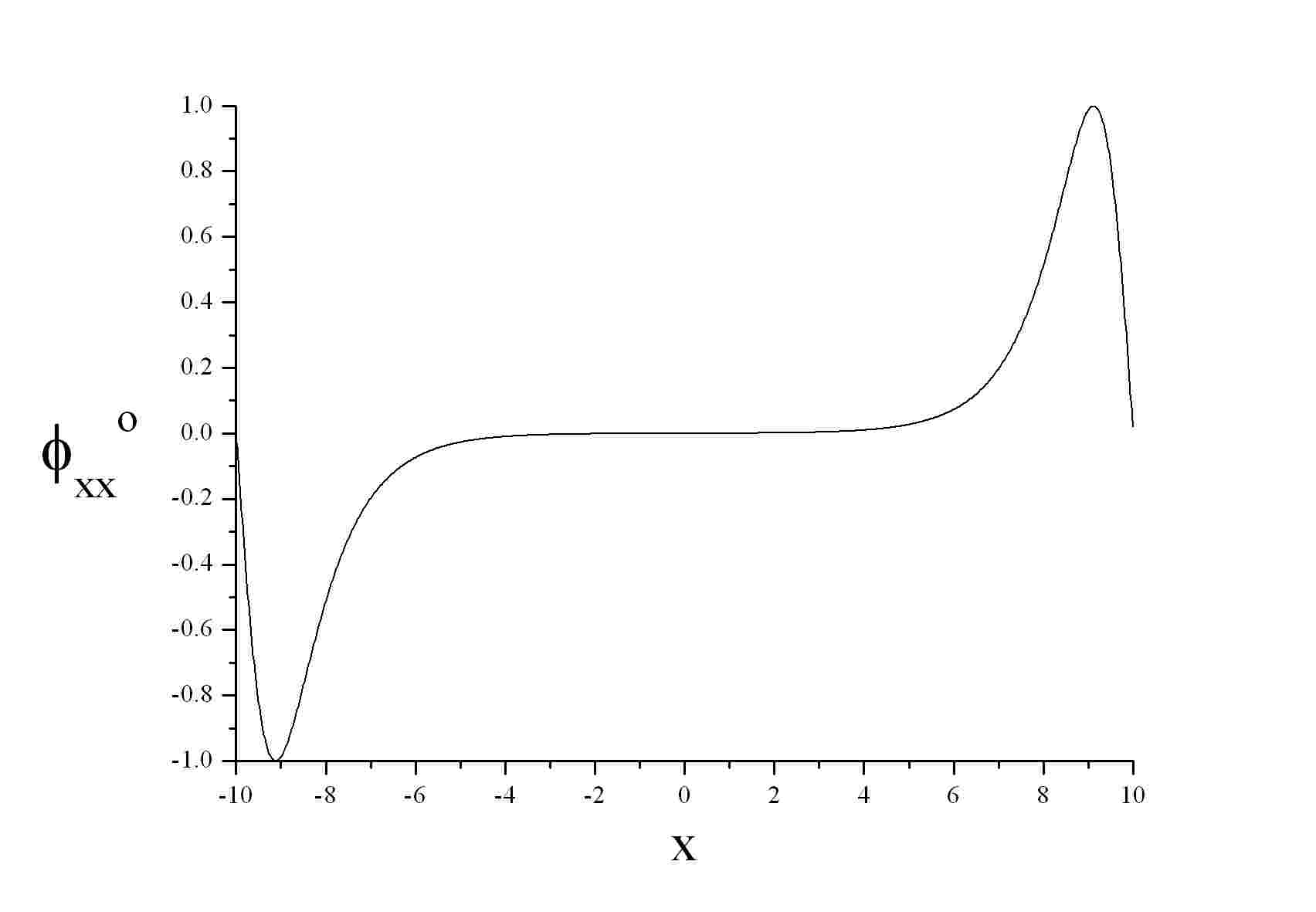}}
\subfigure[ ]{\includegraphics[width=7cm]{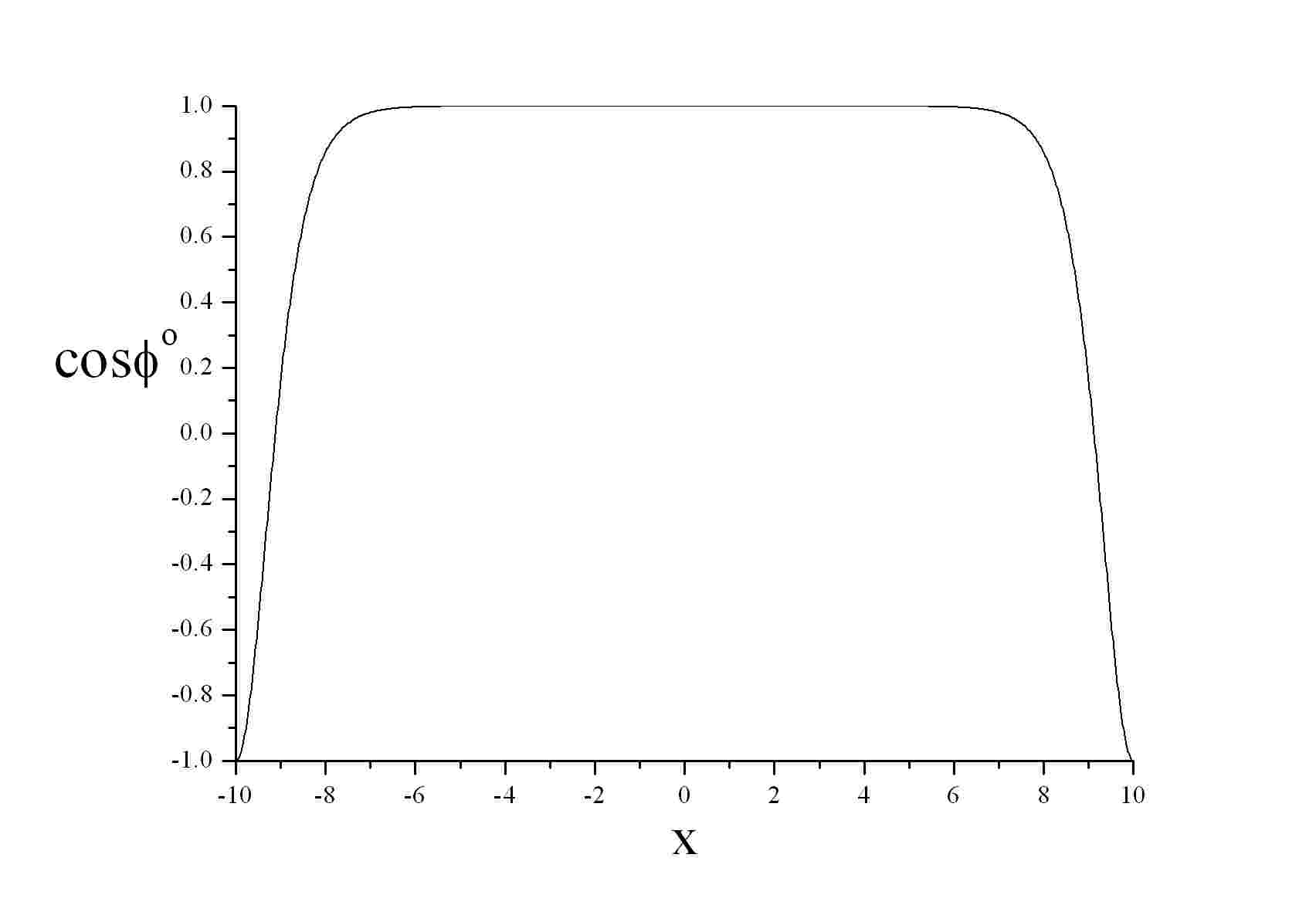}}
\caption{(a) Phase profile $\phi^o(x)$ as in Eq.(\ref{phio2}) for $\gamma=0$, $h=2$ and $l=20$; (b) its first derivative, (c) its second derivative, and (d) its cosines. Note that  $\sin\phi^{o}(x)=\phi_{xx}^{o}(x)$ and $\cos \phi^o(x)= 1-[\phi^o_{x}(x)]^2/2$.}
\label{Figphio}
\end{figure}

\noindent Eq.(\ref{OS}) was first introduced in the analysis of long
asymmetric inline JTJs by Ferrel and Prange\cite{ferrel} in 1963;
however, four years later Owen and Scalapino\cite{OS} reported an
extensive study of its solutions for long symmetric inline JTJs:
this is why Eq.(\ref{OS}) is commonly known as the equation of
Owen-Scalapino(OS). With boundary conditions as in Eq.(\ref{bcn}), the solution of the OS equation has to be
an odd function $\phi^o(x)$ [i.e, with $\phi_x^o(x)$ even]. For
given $h<2$ and $l>\pi/2$, exact solutions exists in terms of Jacobian
elliptic functions. However, upon the assumption that the JTJ is so long that the magnetic field in its center can be neglected, a simple approximate solution exists. In fact, for $l>>1$ (in practice $l>4\pi$), the solution of Eq.(\ref{OS}) with $\phi^o_x(0)=0$ is:

\begin{equation}
\phi^{o}(x)= \pm 4\left[ \tan^{-1} \exp \left(x+\zeta+\frac{l}{2}\right) - \tan^{-1} \exp \left(-x+\zeta+\frac{l}{2}\right) \right],\label{phio2}
\end{equation}

\noindent in which the non-negative constant $\zeta$ is set by
$h=\phi_x^{o}(\pm l/2)\approx \pm 2\textrm{sech} \zeta$ indicating
that the largest possible amplitude of the normalized magnetic
field\cite{likharev} is $h_c=2$ corresponding to $H_c=2 J_c
\lambda_j$ and $\zeta=0$. Now $\phi^o(0)=\pm 4 [\tan^{-1}
\exp(\zeta+l/2) - \tan^{-1} \exp (\zeta+l/2)] = 0$. The sign in
front of Eq.(\ref{phio2}) now concords with that of the applied
field. Eq.(\ref{phio2}) is shown in  Figs.~\ref{Figphio} for $l=20$
and $h=h_c$. Again Eq.(\ref{phio2}), as Eq.(\ref{phie}), represents
a superposition of two static fractional vortices, but now they a
pinned at the junctions extremities and not in its middle point. As
$|h|$ exceeds $h_c$, i.e. $\zeta<0$, we exit the Meissner regime and
some magnetic flux enters into the junction interior; one (or more)
integer vortices (fluxons) gradually develops at each extremity and
move towards the center resulting in a phase profile that can no
more be written in terms of Jacobian elliptic function. For $h>>h_c$
the phase profile resulting from the superposition of several
closely packed fluxons will be approximately linear $\phi^o(x)= hx$
and we recover the behavior of small JTJs (Fraunhofer regime).

\noindent The gradual crossover from intermediate to long JTJs has
to be calculated numerically; here we anticipate that it was found
to be nicely described by the following empirical relationship:

$$h_c(l)=\frac{2 \pi}{l}+ 2 \tanh \frac{l}{2 \pi},$$

\noindent in which the first (Fraunhofer) term dominates for small
values of $l$, while the second (saturating) term dominates for large
$l$ values. It is worth to remark that if no current feeds a junction, then its electrode
configuration does not affect the phase profile; in other words,
inline, overlap and $\delta$-biased JTJs all have the same phase profile as in Eq.(\ref{phio2}) in presence of a given external magnetic field $h$.

\subsection{  $h\neq 0$, $\gamma \neq 0$, $l/4>2\pi$}

\noindent Approximate static phase profiles for non-zero $h$ and
$\gamma$ values are obtained observing that Eq.(\ref{PDE}) can be
rewritten as two identical, although independent, OS PDEs for the
left ($-l/2\leq x < 0$) and right ($0 < x \leq l/2$) inline
asymmetric half-junctions each with an effective bias current
$\gamma/2$. The boundary conditions for the left and right half
junctions are simply given by Eqs.(\ref{bcn}) and (\ref{bc0}):

\begin{eqnarray}
{\left. \frac{d \phi}{d x} \right|_{x=-l/2}} =  & h  \qquad \qquad & {\left. \frac{d \phi}{d x} \right|_{x=0}}  =   \frac{\gamma}{2}  \label{left} \\
{\left. \frac{d \phi}{d x} \right|_{x=0}} =  &  -\frac{\gamma}{2} \qquad \qquad & {\left. \frac{d \phi}{d x} \right|_{x=l/2}}  = h.  \label{right}
\end{eqnarray}

\noindent  It is convenient to rewrite the constant $C$ in
Eq.(\ref{energy}) as $C=(2-k^2)/k^2$; for $C>1$, then $k^2<1$ and
vice-versa. According to the notation of Ref.\cite{OS}, the boundary conditions in Eqs.(\ref{left}) and (\ref{right}) can be rewritten in terms of
Jacobian elliptic functions dn$(x,k_d^2)$ and cn$(x,1/k_c^2)$ of
argument $x$ and modulus $k_d^2$ and $1/k_c^2$, respectively. For
$k_d\leq 1$:

\begin{eqnarray}
\textrm{dn}\left( \frac{l/2-x_{0d}}{k_d} , k_d^2 \right)= \frac{k_d}{2} h \qquad \textrm{and} \qquad
\textrm{dn}\left( \frac{-x_{0d}}{k_d} , k_d^2 \right)= \pm \frac{k_d}{2} \frac{\gamma}{2}  , \nonumber
\end{eqnarray}

\noindent while for $k_c\geq1$:

\begin{eqnarray}
\textrm{cn} \left( {l/2-x_{0c}} , 1/k_c^2 \right)= \frac{k_c}{2} h \qquad \textrm{and} \qquad
\textrm{cn} \left( {-x_{0c}} , 1/k_c^2 \right)= \pm \frac{k_c}{2} \frac{\gamma}{2} . \nonumber
\end{eqnarray}

\noindent Being the above mentioned elliptic functions limited to
the $[-1,1]$ range, the solutions of the OS problem can be found as
far as both $|h|$ and $|{\gamma}/{2}|$ are smaller than $2$. Once
$h\in[-2,2]$, $\gamma\in[-4,4]$ and $l>\pi$ are given, the couples $(x_{0d},k_d)$ or $(x_{0c},k_c)$ can be
numerically found. This mathematical procedure allows to find many
(sometime physically non-interesting) solutions. It is well known
that the number of possible solutions increases with the junction
normalized length\cite{note}.

\begin{figure}[tb]
\centering
\subfigure[ ]{\includegraphics[width=7cm]{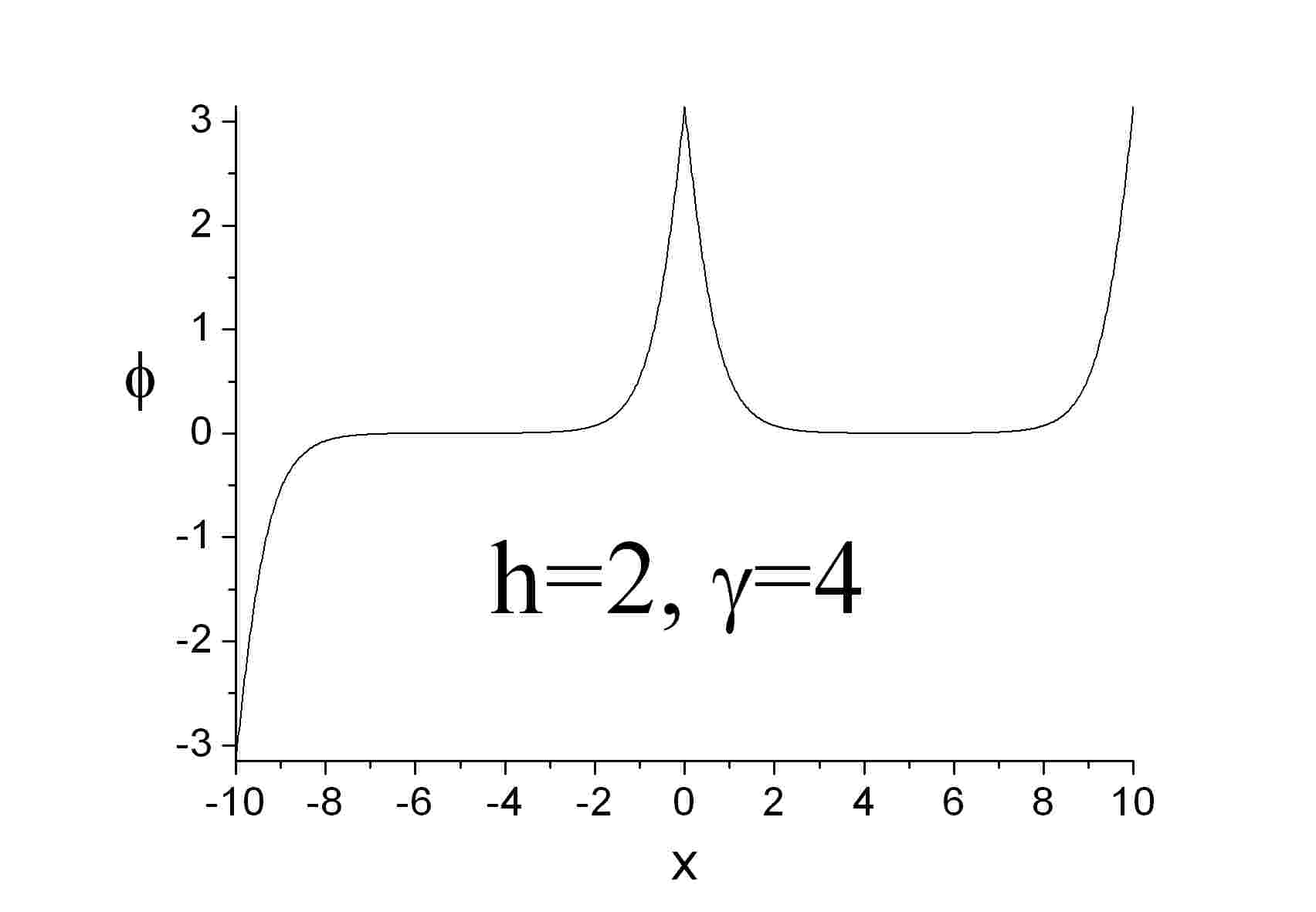}}
\subfigure[ ]{\includegraphics[width=7cm]{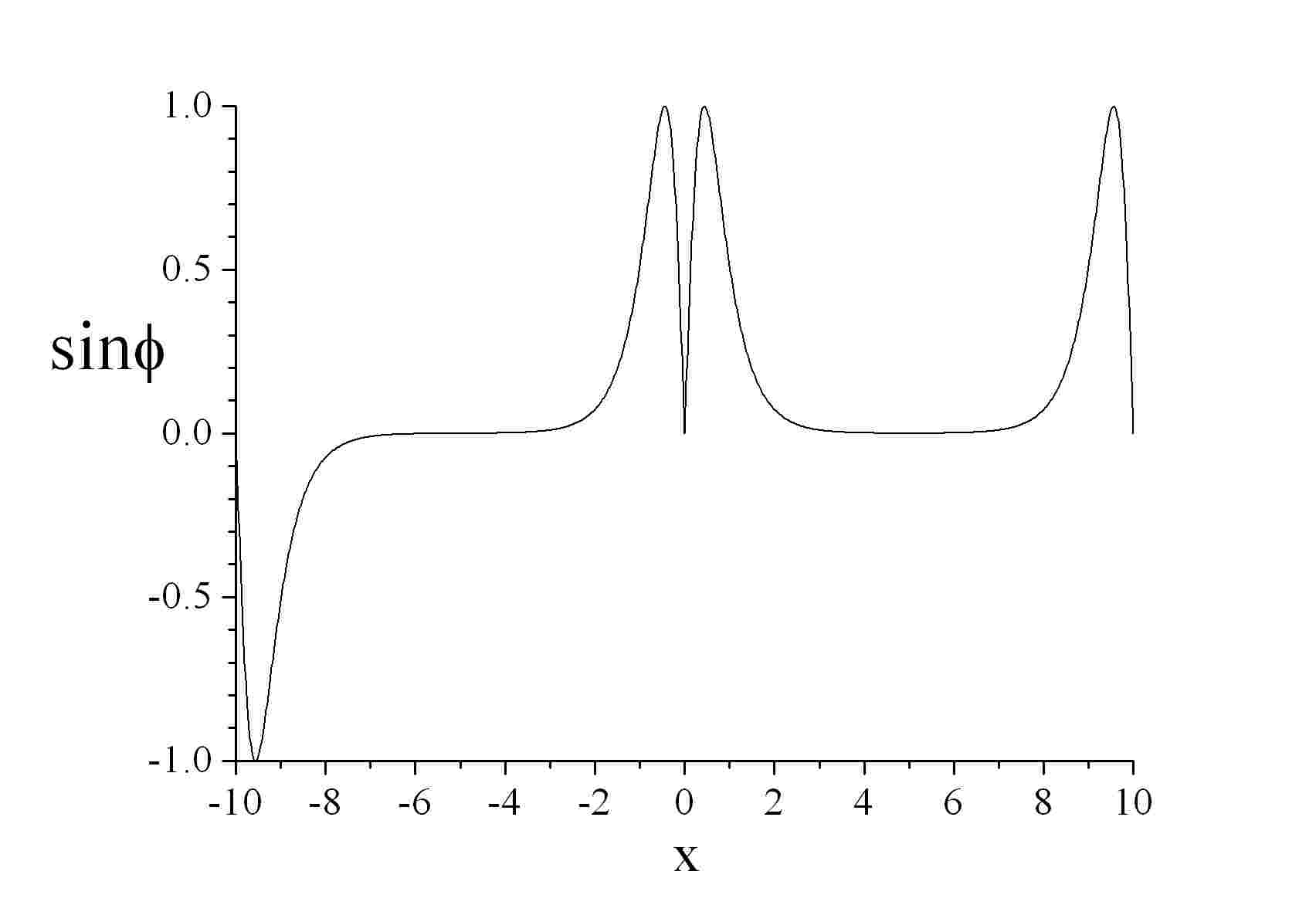}}
\caption{(a) $\phi(x)$ as in (\ref{phi3}) for $h=\gamma/2=2$ ($\zeta= \xi=0$) and $l=20$; (b) $\sin \phi(x)$. Only the junction right side contributes to the Josephson current. It is  $\sin\phi(x)=\phi_{xx}(x)$ and $\cos \phi= 1-[\phi(x)]^2/2$.}
\label{Ling4h2l20}
\end{figure}

\noindent For $l/2>>4\pi$, the approximate solution of Eq.(\ref{OS})
is:

\begin{eqnarray}
\phi(x) &=& \pm 4\left[ \tan^{-1} \exp \left(x-\xi \right) - \tan^{-1} \exp \left(-x+\zeta-\frac{l}{2}\right) \right] \quad \qquad  \textrm{for} \quad -l/2 \leq x < 0,\nonumber\\
&=& \pm 4\left[ \tan^{-1} \exp \left(-x-\xi\right) + \tan^{-1} \exp \left(x-\zeta-\frac{l}{2}\right) \right] \quad \qquad  \textrm{for} \quad 0 < x \leq l/2
\label{phi3}
\end{eqnarray}

\noindent in which $\xi$ and $\zeta$ are two non-negative
independent constants; in fact, $\gamma \approx \pm 4\textrm{sech}
\xi $ and  $h \approx \pm 2\textrm{sech} \zeta $. If $h=0$, then
$\zeta \to \infty$ and $\phi(x)=\phi^e(x)$. Vice-versa, if
$\gamma=0$, then $\xi \to \infty$ and $\phi(x)=\phi^o(x)$. In other
words, the generic static phase profile of a long $\delta$-biased
junction in the Meissner state is obtained simply by the sum of
(four) non-interacting fractional vortices. The sign in front of
Eq.(\ref{phi3}) concords with that of the product $h\gamma$. Looking
at the Eqs.(\ref{left}) and (\ref{right}), it is seen that an
interesting situation occurs when $h=\pm \gamma/2$. For $h=0$ the
bias current $\gamma$ flows symmetrically in the left and right
junction sides [see Fig.~\ref{figPhie}(d)]. With $h\neq 0$ the
symmetry is broken and the current flows mainly in one of the
junction sides. When $|h|=\gamma/2 \leq 2$, then $\zeta=\xi$ and the
applied current only flows in one junction side, while in the other
side the average Josephson current vanishes. For $h= \gamma/2=2$
($\zeta= \xi=0$) the expression above for $\phi(x)$ and its sine are depicted in
Figs.~\ref{Ling4h2l20}(a) and (b), respectively. Only the junction
right side contributes to the Josephson current.

\noindent Let us remark that, being $|\gamma| \leq \gamma_c$, in
force of Eq.(\ref{lineare}), for the largest Josephson current $i_c$
we have:

\begin{equation}
|i_c(h)| = \frac{\gamma_c}{l}. \label{ij8}
\end{equation}

\noindent In other words, for long $\delta$-biased JTJs in the
Meissner state ($-2 \leq h \leq 2$) the critical current is
independent on the externally applied field. There is no other
junction configuration for which this peculiarity occurs. The result
above is only apparently in contrast with Kuprianov\cite{kuprianov}
quadratic prediction for infinitely long point injected junctions:

\begin{equation}
i_c(h) = \frac{2+\sqrt{4+h^2}}{l}, \label{kuprian}
\end{equation}

\noindent leading to $i_c(\pm h_c) = (1+\sqrt{2})i_c(0)/2 \approx
1.21 i_c(0)$. In fact, Eq.(\ref{kuprian}) also takes in the account
the penetration of fluxons into the barrier leading to the
non-Meissner regime characterized by higher modes phase
profiles\cite{OS} that will be discussed in the next Section. In
1985 Radparvar and Nordman\cite{radparvar} reported the experimental
magnetic diffraction pattern of a very long ($L\simeq 100
\lambda_j$) Nb-Pb laterally injected JTJ in reasonable agreement
with Eq.(\ref{kuprian}) despite their electrode configuration only
roughly realized the point injected approximation.

\section{  Static numerical simulations}

\begin{figure}[tb]
\centering \subfigure[ ]{\includegraphics[width=6cm]{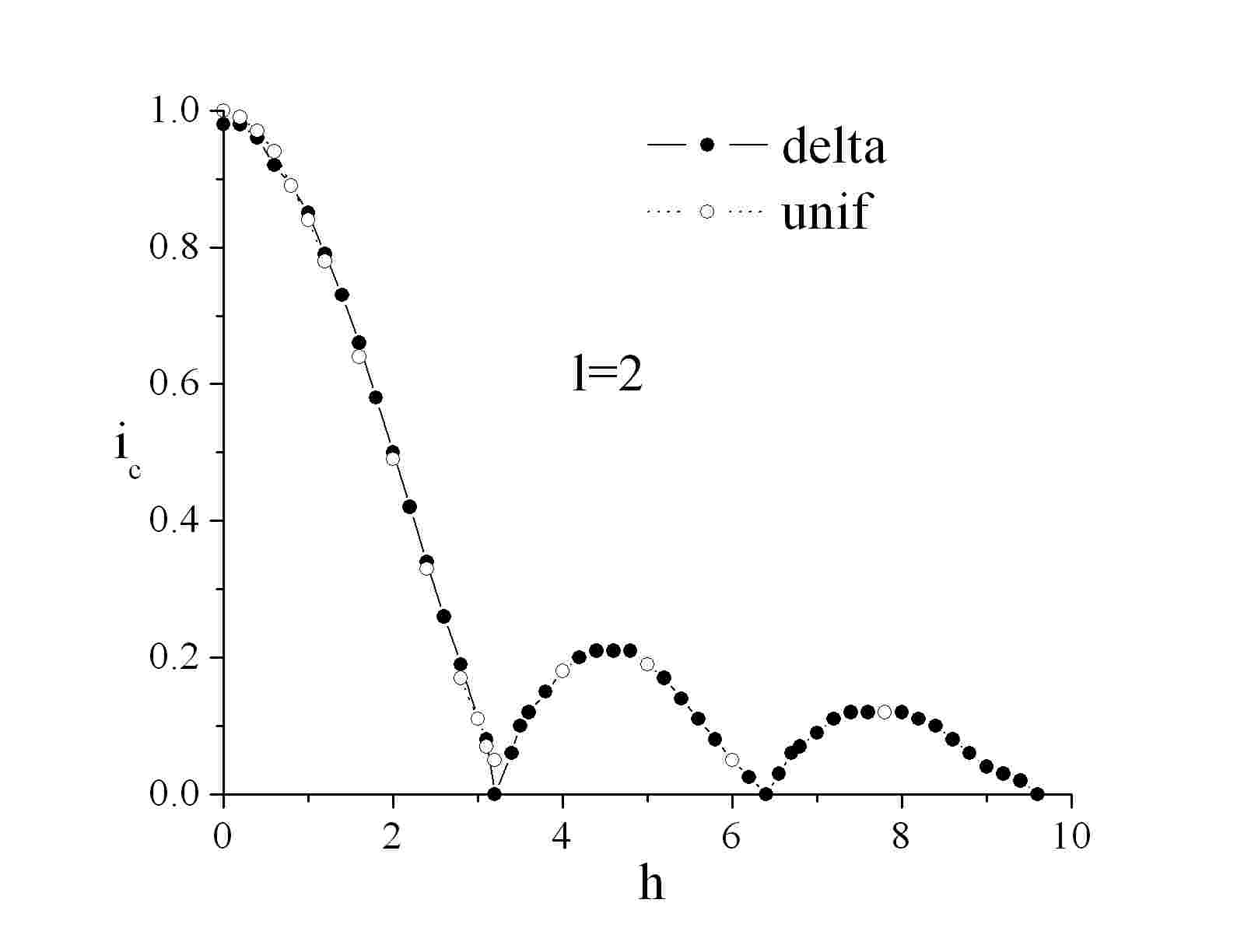}}
\subfigure[ ]{\includegraphics[width=6cm]{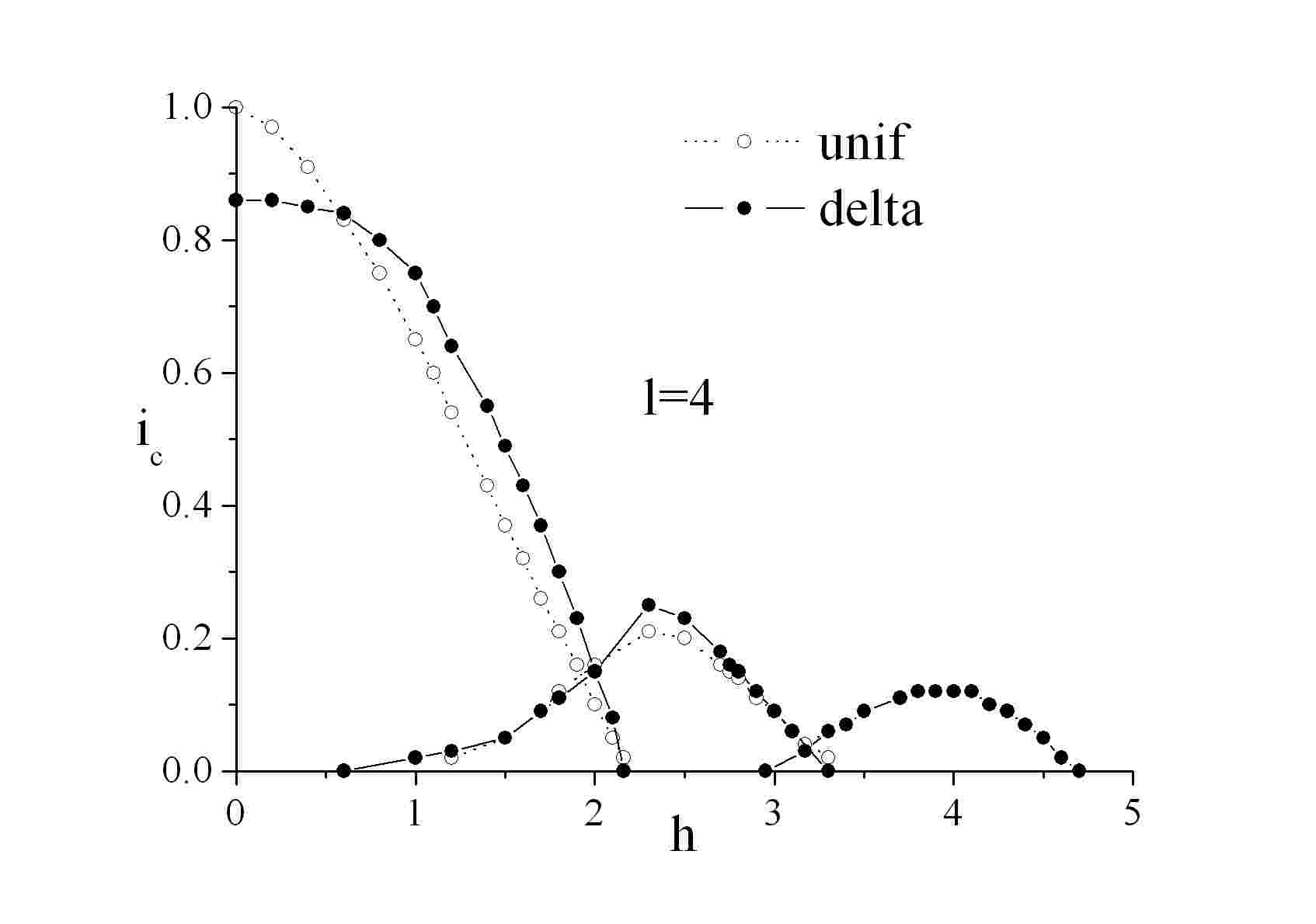}} \subfigure[
]{\includegraphics[width=6cm]{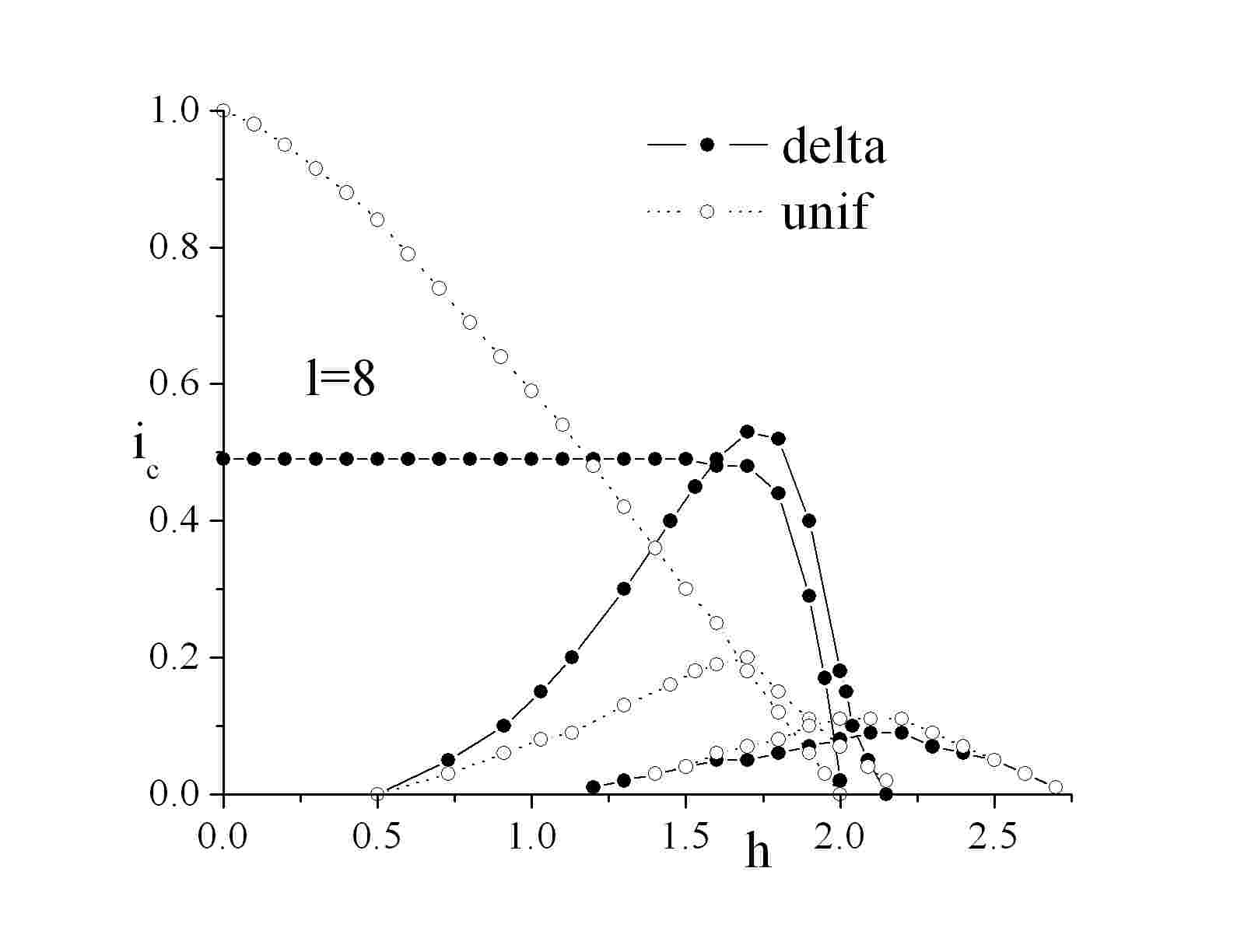}} \subfigure[
]{\includegraphics[width=6cm]{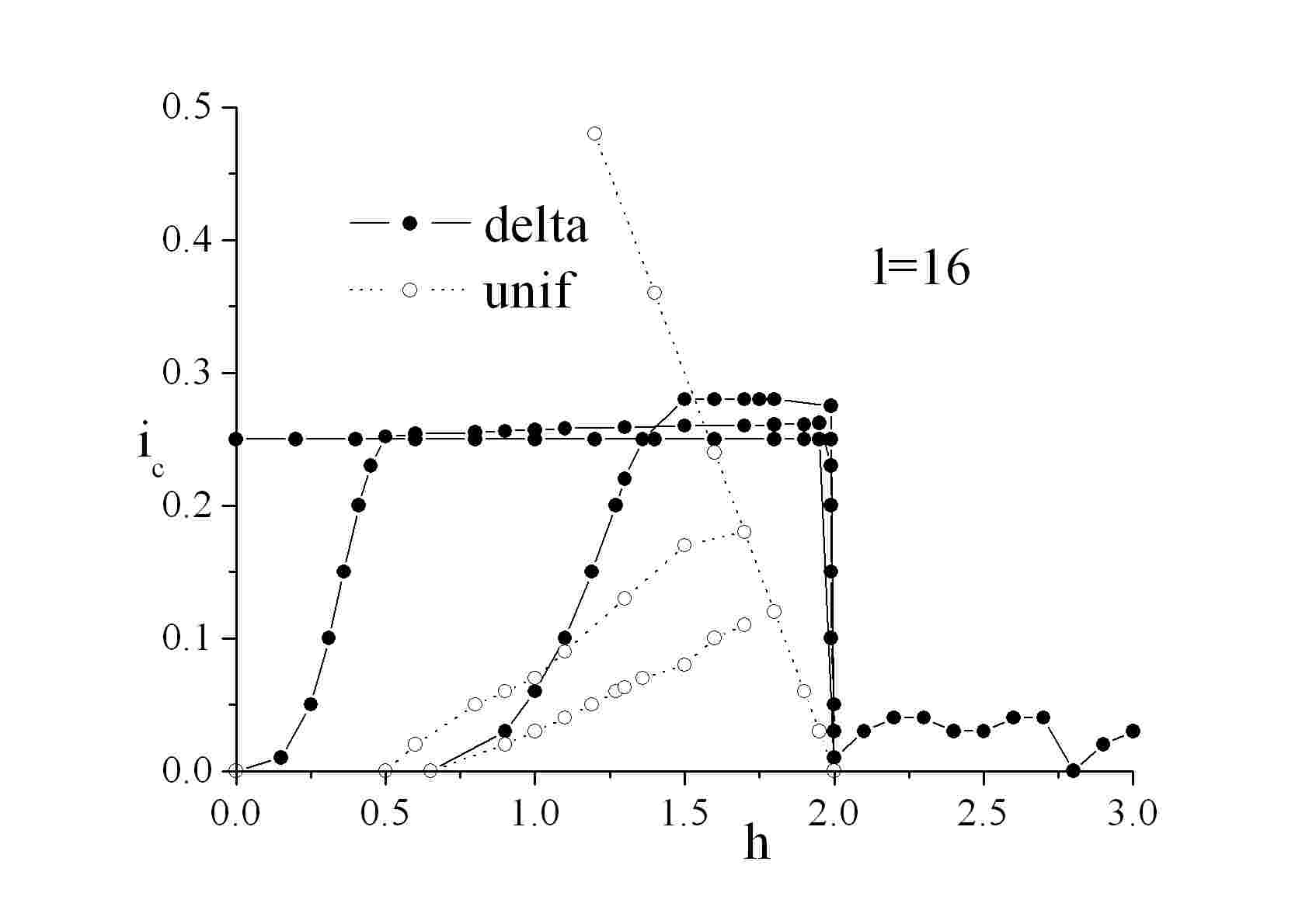}} \caption{Numerically
computed magnetic diffraction patterns for for different junction
normalized lengths. From (a) to (d) $i_c$ vs. $h$ respectively, for
$l=2,4,8,16$. The full dots refer to point injected bias current,
while, for the sake of comparison, the open dots correspond to the
well-known case of a uniformly biased junction.} \label{linear}
\end{figure}

\noindent In this section we discuss the numerically obtained
solutions of the PDE in Eq.(\ref{PDE}) with boundary conditions as
in Eq.(\ref{bcn}). The direct numerical integration of Eq.(\ref{PDE})
poses large problems of stability due to the fact that there are no
losses in the system; to avoid this problems, we recurred to the
integration of the time dependent perturbed sine-Gordon equation:

\begin{equation}
 \phi_{xx} - \phi _{tt}-\sin \phi = \gamma \delta(x)
+ \alpha \phi_t - \beta \phi_{xxt},
\label{time}
\end{equation}

\noindent with $\alpha=3$ in order to have a fast decay of the
temporal features of the solution towards a static solution (in real
device $\alpha \le 0.01$). The term containing the surface losses
was simply dropped to save computer time, i.e. $\beta=0$.
Eq.(\ref{time}) has been numerically integrated by using the
commercial finite-element simulation package of Comsol Mutiphysics
(www.comsol.com) for different values of the normalized length $l$
which enters the PDE through the boundary conditions. The
$\delta$-function has been approximated by the continuous function
$f(x)= \rho \textrm{sech}^2(2\rho x)$ with the parameter $\rho=O(10^2)$ [note that
$f(0)=\rho$, $f(1/\rho)=\rho\textrm{sech}^2 2\approx  0.07\rho$ and
$\int_{-\infty}^{\infty} f(x)dx=1$].

\noindent In order to trace the different lobes of $i_c$ vs. $h$, it
is crucial to start the numerical integration with a proper initial
phase profile. As far as $l<\pi$ the initial condition for the
numerical integration of Eq.(\ref{PDE}) was simply set to
$\phi(x)=0$. In order to find the several possible initial
conditions for $l>\pi$, the OS problem was solved for a given value
of $h$ and with $\gamma$ set to zero. Once the initial phase
derivative was known, also the initial phase profile could be easily
derived\cite{OS}. Finally, during the numeric calculation, $\gamma$
was changed until the numerical solution becomes unstable. We have
numerically computed $\gamma_c$ as the maximum allowed value of the
bias current for each chosen value of the magnetic field $h$. Once
$\gamma_c$ was found, the corresponding critical current $i_c$ could
be calculated either numerically or, in a equivalent manner,
resorting to Eq.(\ref{lineare}). As expected, a general peculiarity
of the numerically found phase profiles for long enough
$\delta$-biased junctions is that the phase values at the junction
extremities ($x=\pm l/2$) do not depend on $\gamma$; viceversa, the
phase behavior near the origin ($x=0$) does not depend on the
applied field $h$.

\noindent Figs.~\ref{linear}(a)-(d) display the magnetic diffraction
patterns for point injected JTJs with different normalized lengths.
We remark that the $i_c(h)$ patterns are symmetric around $h=0$. For
comparison we also report the $i_c(h)$ for a uniform bias; more
precisely, the full dots refer to the point injected current, while
the open (gray) dots correspond to the uniform bias. For $l=1$, the
numerical data (not reported) closely follow the expected
Fraunhofer-like dependence (with $h_c=2\pi$) with differences only
in the third significant digit. As shown in Fig.~\ref{linear}(a),
for $l=2$ we still have a Fraunhofer-like pattern. Pronounced
deviations from small \jun behavior were found for $l=4$, but, as
can be seen in Fig.~\ref{linear}(b), they disappear for large filed
values. Increasing $l$, some ranges of magnetic field develop in
correspondence of the pattern minima in which $i_c$ may assume two
(or more) different values corresponding to different phase profiles
inside the barrier. In fact, each pattern lobe is associated with a
given vortex structure; more precisely, in the first lobe which, for
$l=4$, goes from $h=0$ to $h_c\approx 2.2$, the external magnetic
field is shielded and vortex cannot penetrate into the barrier
(Meissner state). However, at the very end of this lobe a  fluxon is
present in the junction. In the successive lobes the magnetic
field penetrates into the barrier and vortices are created in the
barrier in a way which closely recalls the behavior of the type II
superconductors, even though the vortices we are dealing with are
quite different from the Abrikosov ones as they do not have a
normal core. In the second lobe, moving from $h\approx 3$ to $h
\approx 4.6$, we start from a phase configuration very similar to
that at the right side of the first lobe in which one vortex is
present in the barrier and we end up with two bunched fluxons.
Adopting the terminology used in Ref.\cite{OS}, we refer to the
first (Meissner) lobe as to '$0$ to $1$ vortex mode' lobe, the
second as the '$1$ to $2$ vortex mode' lobe and so on. In general,
one may talk about the '$n$ to $n+1$ vortex mode' when the junction
contains more than $n$ but less than $n+1$ vortices. As $l$ is
increased from $4$ to $8$ drastic changes occur (the crossover point
being approximately $2\pi$), as shown in Fig.~\ref{linear}(c): the
critical current get smaller and smaller, according to
Eq.(\ref{ic}), and the principal lobe gets flatter and flatter
resulting in a rather large plateau; further, the lobes broaden and
overlap each other with $h_c$ converging to $2$. This behavior is
even more pronounced for $l=16$ as shown in Fig.~\ref{linear}(d). In
other words, the main (Meissner) lobe is a rectangle with corners
in $h=0$ and $2$ and $\gamma=0$ and $4$. The phase profiles in the
non-trivial corners are those already shown in Figs.~\ref{figPhie},
\ref{Figphio} and \ref{Ling4h2l20}.

\begin{figure}[tb]
\centering
\subfigure[ ]{\includegraphics[width=7cm]{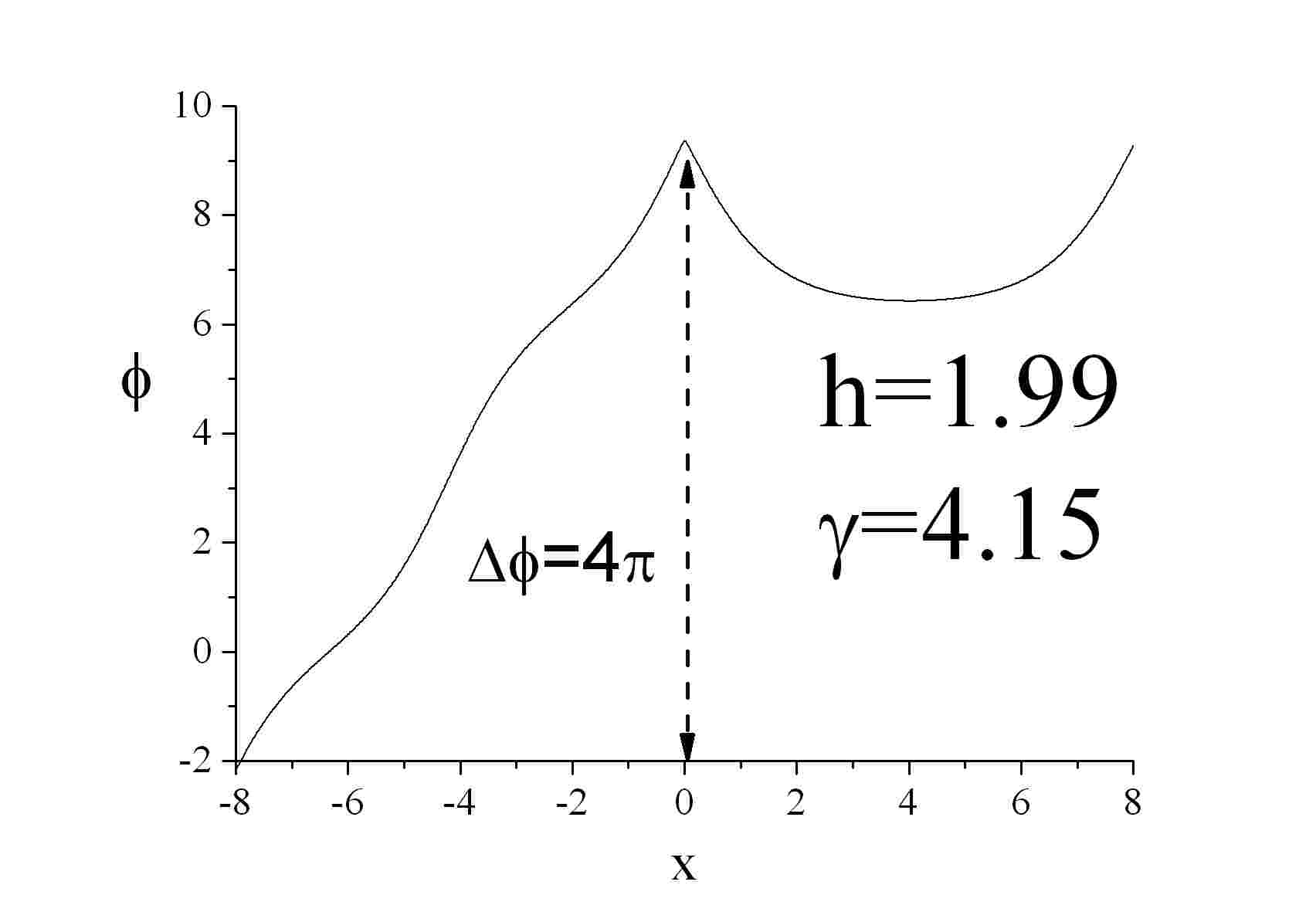}}
\subfigure[ ]{\includegraphics[width=7cm]{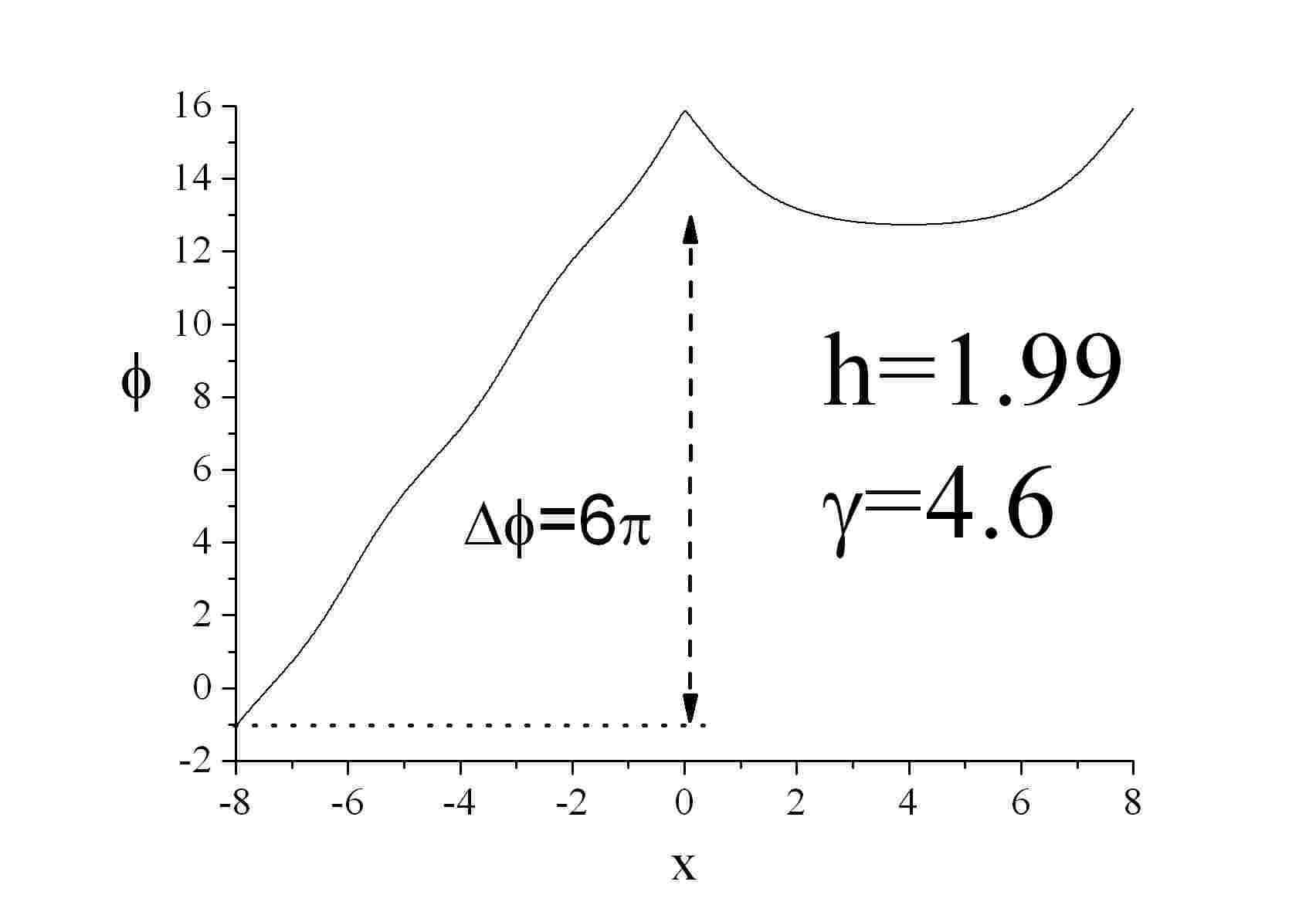}}
\caption{Numerically computed phase profiles for higher order modes with $l=16$ and $h=1.99$ and different $\gamma$ values: (a) 2 vortex mode for $\gamma=4.15$ and (b) 3 vortex mode for $\gamma=4.6$.}
\label{profiles}
\end{figure}

\noindent The phase profile corresponding to the 2 and 3 fluxon
modes are reported in Figs.~\ref{profiles}(a) and (b), respectively,
for $h$ ($\gamma$) slightly below (above) its critical value. It is
evident that the vortices only penetrate in left side of the
junction. However, the situation reverses by reversing either $h$ or
$\gamma$. The maximum supercurrent associated to a given vortex mode
increases with the mode order; this is peculiar of $\delta$-biased
JTJs.

\section{  Experiments}

\begin{figure}[b]
        \centering
                \includegraphics[width=8cm]{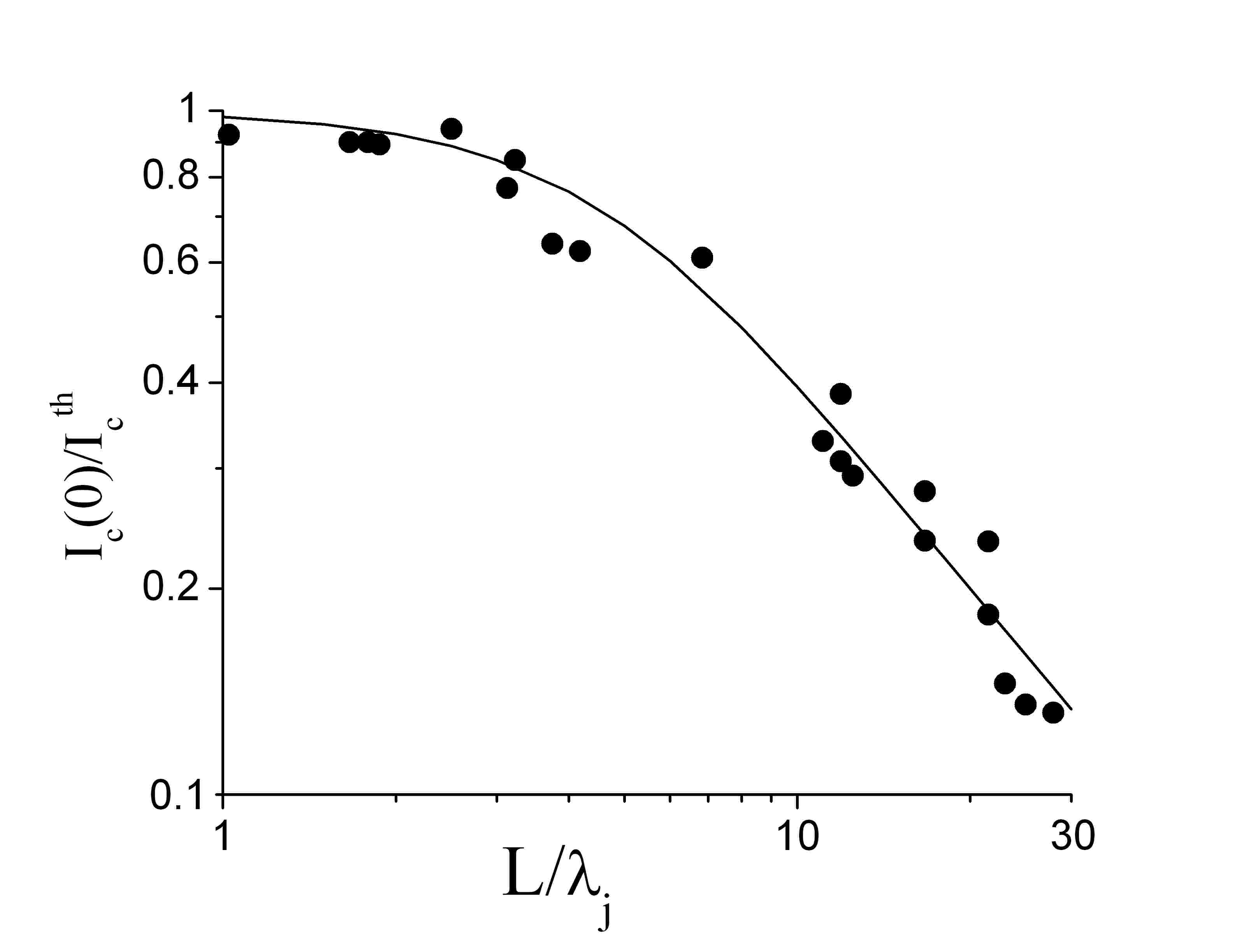}
        \caption{log-log plot of the measured zero-field critical currents $i_c(0)$ of many $\delta$-biased JTJs versus their reduced length $L/\lambda_j$. The critical current has been normalized to the small junction theoretical value calculated as the $70\%$ of the current jump at the junction gap voltage. The solid line is obtained from Eq.(\ref{ic}) with $\gamma_c$ as in the empirical Eq.(\ref{gammac}).}
        \label{IcData}
\end{figure}

To investigated the properties of $\delta$-biased JTJs, we have used
are high quality $Nb/Al-Al_{ox}/Nb$ JTJs fabricated on silicon
substrates using the trilayer technique in which the junction is
realized in the window opened in a $SiO$ insulator layer. We
measured a large number of both linear and gapped annular junctions
all of them with width $W=2\,\mu$m, but whose lengths varied from a
$L=80$ to $335\,\mu$m. The width of the electrodes carrying the
current in and out of the barrier was $10\,\mu$m. The thickness of
the $SiO_2$ insulator layer was $200\,nm$ and the so called ''idle
region'', i.e. the overlapping of the wiring layer onto the base
electrode was about $1\,\mu$m for all the junctions. In order to
vary the sample normalized lengths over a large range values, we
used two sets of samples having quite different critical current
densities ($J_c=100\,A/cm^2$ and $3kA/cm^2$), corresponding to
$\lambda_j\simeq 80$ and $\simeq 12\,\mu$m. In such a way samples
were available with $l$ spanning from about $1$ to about $30$. The
values of the Josephson penetration depth were calculated taking
into account the the effect of the lateral idle
region\cite{JAP95,Ustinov}. In Fig.~\ref{IcData} we report the
log-log plot of the measured zero-field critical currents $i_c(0)$
of many $\delta$-biased JTJs versus their reduced length
$L/\lambda_j$. The critical current has been normalized to the small
junction theoretical value calculated as the $70\%$ of the current
jump at the junction gap voltage. We observe that, for large
normalized length, the experimental data clearly follow an inverse
proportionality law as expected from Eq.(\ref{ic}) and the
transition from long to short junctions is nicely fitted by the
empirical expression in Eq.(\ref{gammac})(solid line).

\vskip 5pt

\begin{figure}[tb]
\centering
\subfigure[ ]{\includegraphics[width=7cm]{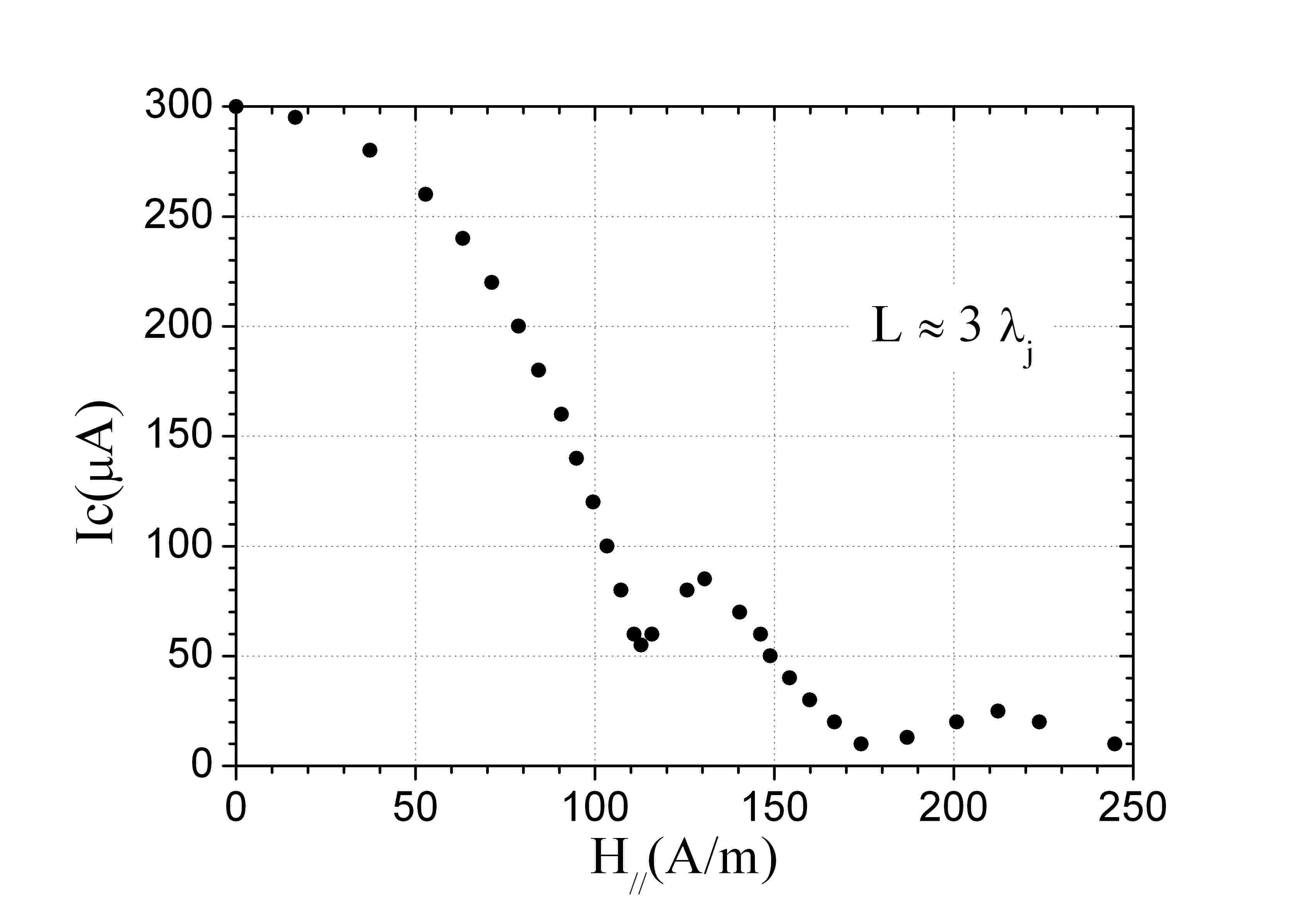}}
\subfigure[ ]{\includegraphics[width=7cm]{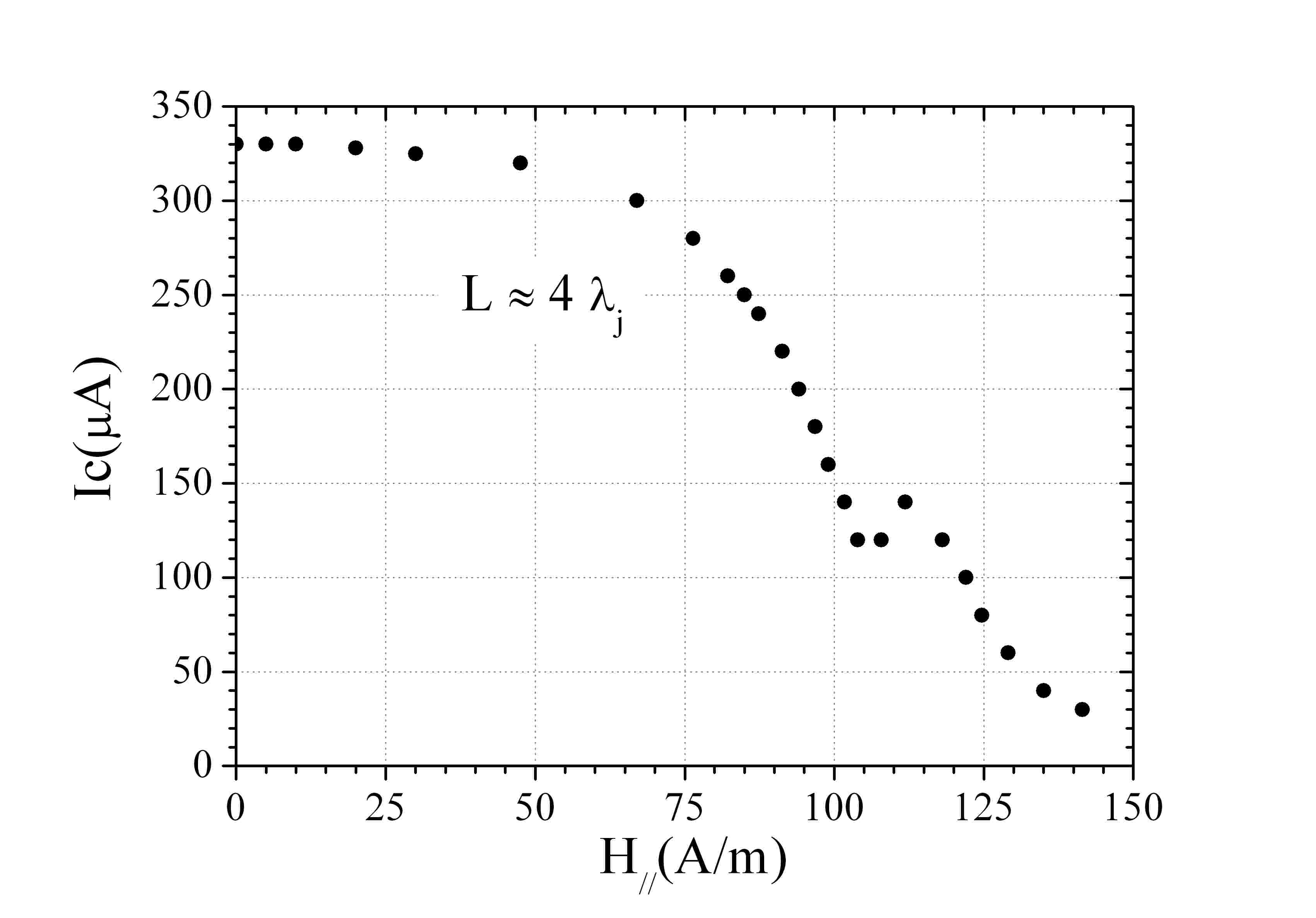}}
\subfigure[ ]{\includegraphics[width=7cm]{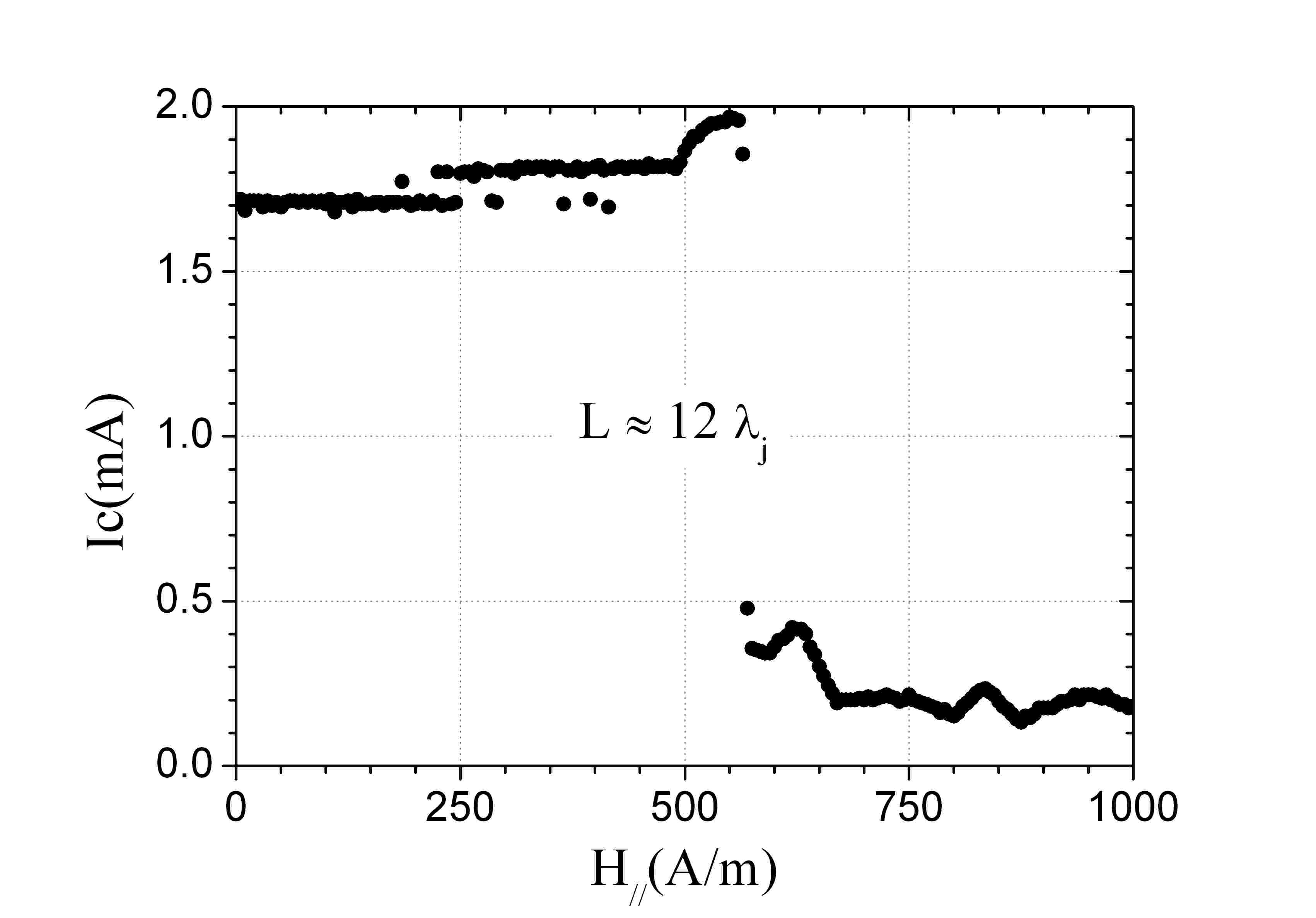}}
\subfigure[ ]{\includegraphics[width=7cm]{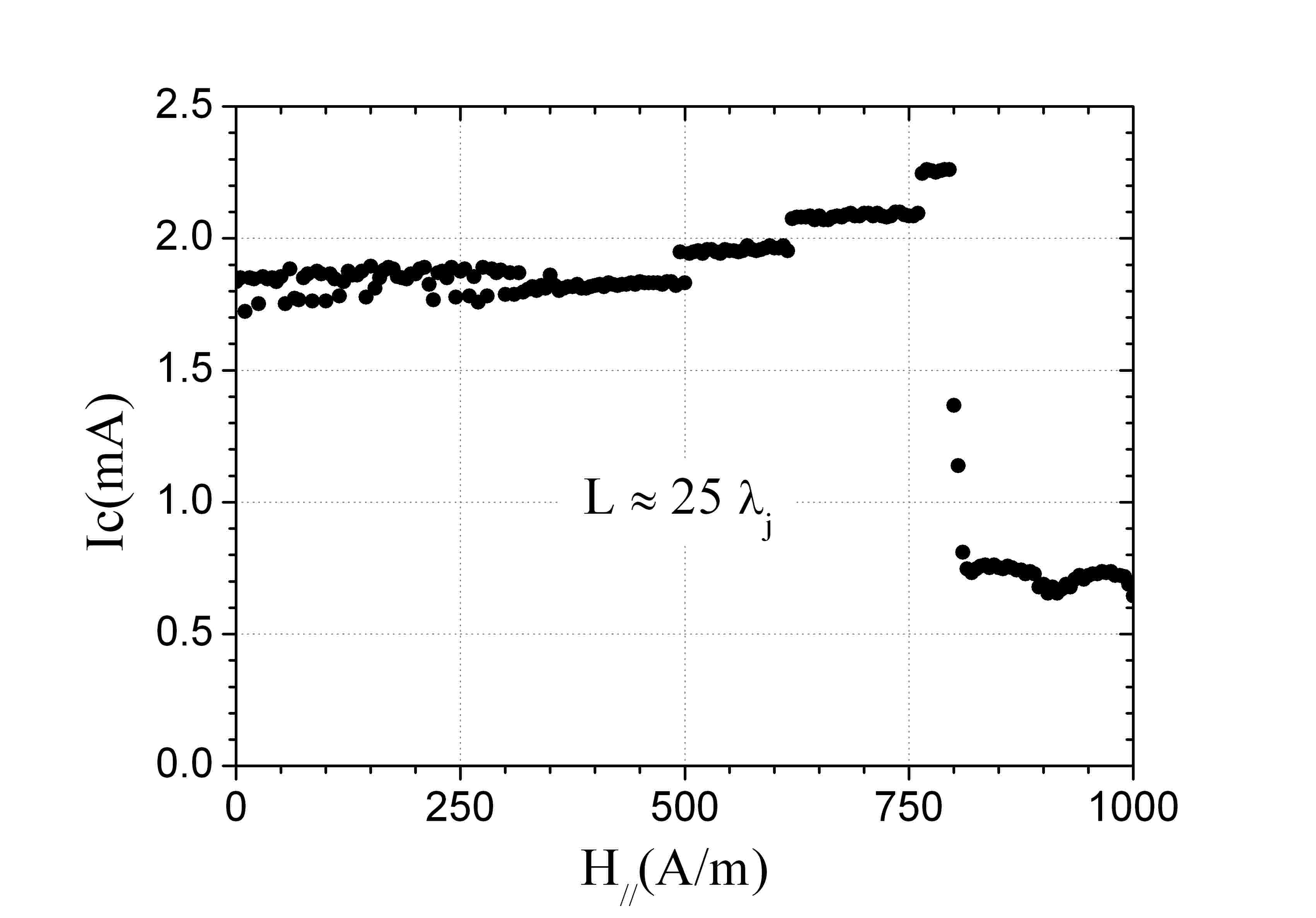}}
\caption{Experimental magnetic diffraction patterns taken for linear samples in a uniform in-plane field with increasing normalized lengths $l=L/\lambda_j$. (a) $L=250\,\mu$m and $\lambda_j\simeq 80\,\mu$m, (b) $L=300\,\mu$m and $\lambda_j\simeq 80\,\mu$m, (c) $L=150\,\mu$m and $\lambda_j\simeq 12\,\mu$m, and (d) $L=300\,\mu$m and $\lambda_j\simeq 12\,\mu$m.} \label{exp}
\end{figure}

\noindent On real samples, the measurements of maximum supercurrent
against the external field often yield the envelop of the lobes,
i.e., the current distribution switches automatically to the mode
which for a given field carries the largest supercurrent. Sometimes,
for a given applied field, multiple solutions are observed on a
statistical basis by sweeping many times on the \jun current-voltage
characteristic. Figs.~\ref{exp}(a)-(d) display the measured magnetic
diffraction patterns for four linear samples in a uniform in-plane
field with selected normalized lengths. Analogous results (not
reported) were obtained for samples with the gapped annular geometry
in an uniform, although uncalibrated, radial magnetic field. We
found an excellent agreement with the results of numerical
calculation discussed in the previous section. Marked deviation from
the small junction Fraunhofer-like behavior were observed for $L=3
\lambda_j$. It is evident that, for longer JTJs, the $I_c(H)$ is
made of few flat segments corresponding to different static phase
solutions containing more and more fluxons. Since the number of
possible static solutions increases with the junction normalized
length, it is straightforward to assume that, as $l$ increases, the
number of flat segments increases too, while their lengths decrease
so that to merge in a monotonically increasing critical current as
predicted in Ref.\cite{kuprianov} and observed in
Ref.\cite{radparvar}.

\begin{figure}[htb]
        \centering
                \includegraphics[width=8cm]{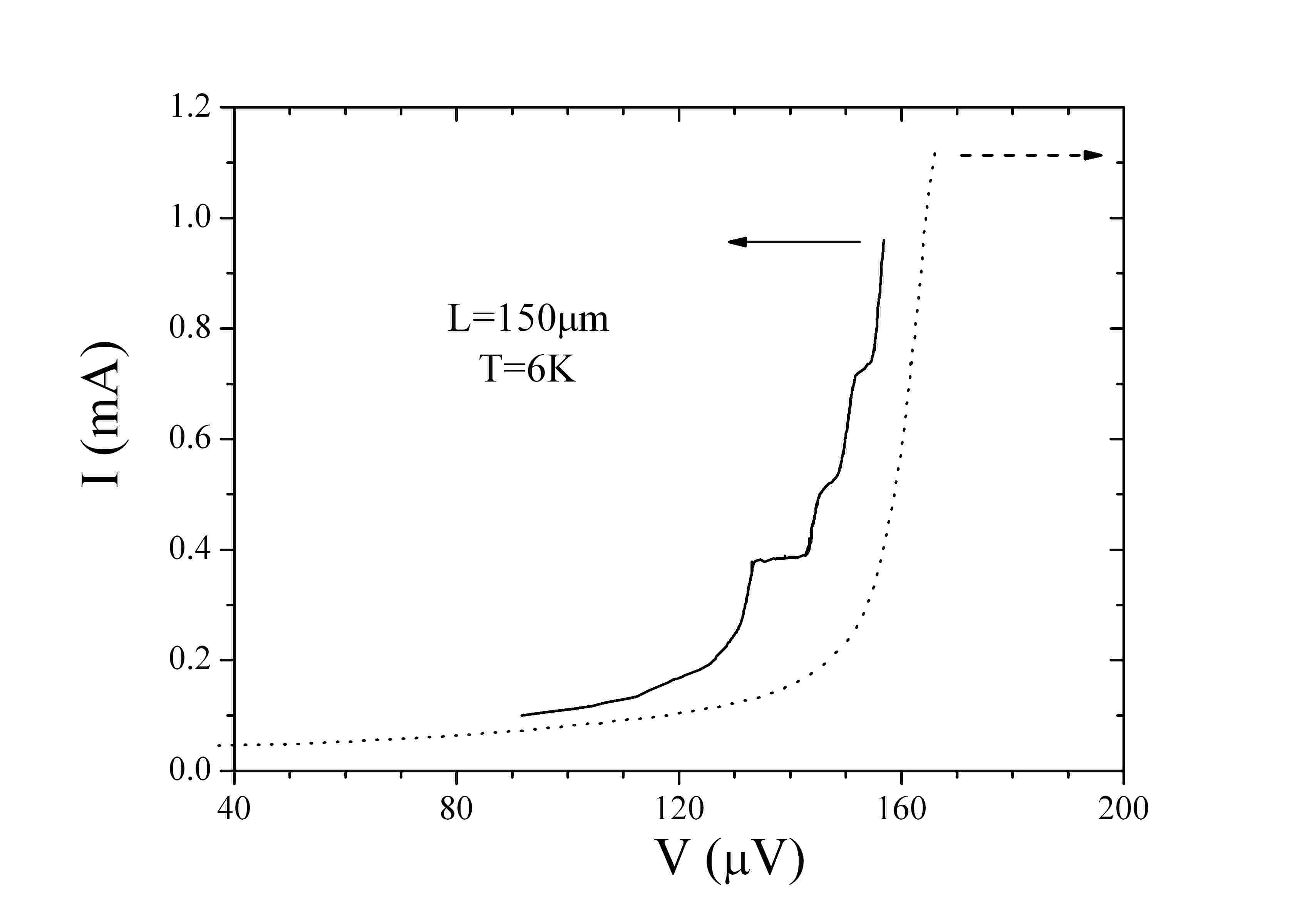}
        \caption{ Profiles of the first zero-field step for overlap junctions with $L=150\,\mu$m and $J_c=3\,kA/cm^2$ (corresponding to $\lambda_j\simeq 12\,\mu$m). (a) $\delta$-bias (solid line) and (b) uniform bias (dashed line). The data were taken a $T\approx 6K$. The arrows indicate the switching direction at the top of the step.}
        \label{zfs}
\end{figure}

\vskip 5pt

\noindent The dynamic of a $\delta$-biased JTJ was also investigated
both experimentally and by numerically integrating the time
dependent Eq.(\ref{time}) whose steady state solutions were typical
of a soliton moving in a two-level potential, with the fluxon(s)
being accelerated only when passing trough the origin. We have
focused out attention on the zero-field single fluxon shuttling
mode. Figs.~\ref{zfs} compares the first zero-field step profiles of
a $\delta$-biased (solid line)and a uniformly biased (dashed line)
long overlap JTJs having the same length ($L=150\,\mu$m) and the
same Josephson current density ($J_c=3\,kA/cm^2$). The data were
taken a $T\approx 6K$. Contrary to the case of uniform bias, a
current threshold value is needed to avoid that fluxon stops due to
friction. This value, of course, depends on the the junction length
and losses: the longer the junction, the smaller should the losses
be to maintain a finite-voltage dynamic state. The
$\gamma$-$\left\langle \phi_t \right\rangle$ plot, corresponding to
the zero-field step profile, is smoother that that obtained when the
bias is uniform and is characterized by fine structures associated
to the fluxon interactions with plasma waves\cite{golubov,barbara}
originated by the fluxon itself when passing across the potential
discontinuity. Often the back-switching transition has been observed
at the top of the step, as indicated by the solid arrow pointing to the zero
voltage state. As a matter of fact, point injected junctions seem
not to reveal new dynamical states. We only remark that
\textit{displaced linear slopes}\cite{barone} were numerically found
and experimentally regularly observed in absence of magnetic field
and for junction with large losses ($\alpha\approx 1$ or $T >
0.7\,T_c$). when the d.c. bias exceeded its critical value.
Displaced linear slopes were also observed in a presence of a
magnetic field which, by increasing the field further, eventually
develop in large amplitude flux flow steps.

\section{Conclusions}

\noindent In this paper we focused on $\delta$-biased (or
point-injected) long Josephson tunnel junctions. Our analysis goes
much beyond the previous theoretical\cite{kuprianov} and
experimental\cite{radparvar} works in which only extremely long
junctions were treated. We considered a linear JTJ in a uniform
magnetic field and a gapped annular junction in a radially uniform
field. These two electrode configurations (shown in
Figs.~\ref{2D}(b) and (a), respectively) are topologically
equivalent, that is, they are described by the same partial
differential equation [Eq.(\ref{sG})] with the same boundary
condition [Eq.(\ref{bcn})]. Apart from their intriguing physical
properties, the interest for $\delta$-biased JTJs stems from the
fact that they were successfully used to detect trapped fluxoids in
a recent experiment aimed to study the spontaneous defect production
during the fast quenching of a superconducting loop\cite{PRB09}. The
main peculiarity of a $\delta$-biased junction is the jump of the
phase derivative at the point were the current is injected. We have
shown that this discontinuity naturally leads to the formation of
fractional vortices. In the last few years there has been a great
deal of interest in the phase discontinuity observed in $0-\pi$
transition junctions which can be modeled with a bias made of two
closely spaced $\delta$-functions with opposite sign, more precisely
by the derivative of a $\delta$-function\cite{goldobin09}. We believe that our findings on the properties of a {\it single} $\delta$-function can shed some more light on the mechanisms responsible of the appearance of fractional vortices in $0-\pi$ transition Josephson tunnel junctions.

\vskip 5pt

\noindent The authors thank  A. Gordeeva for useful discussions and
for helping us with the dynamical numerical simulations. One of us
(VPK) acknowledges the financial support from the Russian Foundation
for Basic Research under the grants 09-02-00246 and 09-02-12172.


\begin{thebibliography}
\bibitem{note} Let us remind that for $k_c<1$,
cn$(u,1/k_c^2)$ is a periodic odd function with period $U_c=4
K(1/k_c^2)$ and, for $k_d<1$, dn$(u,k_d^2)$ is periodic even with
period $U_d=2 K(k_d^2)$ (where $K(k^2)$ is the complete elliptic
integral of first kind: $k^2\in[0,1]$ and
$K(k^2)\in[\pi/2,\infty]$). In the limit $k_{c,d} \to 1$
(corresponding to the limit $l \to \infty$), the period diverges and $\textrm{cn}(u,1)=\textrm{dn}(u,1)=\textrm{sech} u$.
\end{thebibliography}
\end{document}